\documentclass[aps,pre,twocolumn,showpacs,superscriptaddress,preprintnumbers,ams
math,amssymb]{revtex4}

\usepackage{amsmath}
\usepackage{graphicx}
\usepackage{dcolumn}
\usepackage{bm}
\usepackage{float}
\usepackage{color}
\usepackage{subfig}
\usepackage{textcomp}
\usepackage{ifpdf}

\begin{document}

\author{Q. Moreno}\affiliation{University of Bordeaux, Centre Lasers Intenses et Applications, CNRS, CEA, UMR 5107, F-33405 Talence, France}

\author{M. E. Dieckmann}\affiliation{Department of Science and Technology, Link\"oping University, SE-60174 Norrk\"oping, Sweden}

\author{X. Ribeyre}\affiliation{University of Bordeaux, Centre Lasers Intenses et Applications, CNRS, CEA, UMR 5107, F-33405 Talence, France}

\author{S. Jequier}\affiliation{University of Bordeaux, Centre Lasers Intenses et Applications, CNRS, CEA, UMR 5107, F-33405 Talence, France}

\author{V. T. Tikhonchuk}\affiliation{University of Bordeaux, Centre Lasers Intenses et Applications, CNRS, CEA, UMR 5107, F-33405 Talence, France}

\author{E. d'Humi\`eres}\affiliation{University of Bordeaux, Centre Lasers Intenses et Applications, CNRS, CEA, UMR 5107, F-33405 Talence, France}

\date{\today}

\title{Impact of the electron to ion mass ratio on unstable systems in particle-in-cell simulations}

\begin{abstract}
The evolution of the Buneman and two-stream instabilities driven by a cold dilute mildly relativistic electron beam is studied as a function of the ion-to-electron mass ratio. The growth rates of both instabilities are comparable for the selected parameters if the realistic ion-to-electron mass ratio is used and the Buneman instability outgrows the two-stream instability for an artificially reduced mass ratio. Particle-in-cell (PIC) simulations show that both instabilities grow independently during their linear growth phase. The much lower saturation amplitude of the Buneman instability implies that it saturates first even if the linear growth rates of both instabilities are equal. The electron phase space holes it drives coalesce. Their spatial size increases in time and they start interacting with the two-stream mode, which results in the growth of electrostatic waves over a broad range of wave numbers. A reduced ion-to-electron mass ratio results in increased ion heating and in an increased energy loss of the relativistic electron beam compared to that in a simulation with the correct mass ratio. 
\end{abstract}

\maketitle
\section{Introduction}

A charge- and current neutral collision-less plasma, which is composed of two counterstreaming electron beams and ions at rest, is unstable. Such systems are frequently found in space- and astrophysical plasmas as well as in laboratory plasmas. They develop if a fast electron beam enters an initially unperturbed plasma at rest \cite{Lovelace71,Thode75}. The net current carried by the electron beam drives electromagnetic fields, which accelerate the electrons of the background plasma into the opposite direction. The return current of the latter eventually balances that of the beam, which restores the plasma's current-neutrality.

Several instabilities can develop after current-neutrality has been reestablished. The two-stream instability, which is driven by the interaction of both electron beams, competes with one of two possible electrostatic instabilities if we constrain the wave vector to the direction along which the beams are drifting. 

It competes with the Buneman instability if the drift speed between the background electrons and the ions exceeds significantly the thermal speed of both species. The Buneman instability originally refers to the instability of one electron beam with one ion beam with the same charge density \cite{Buneman58}. Here we use the term Buneman instability to denote an instability between an electron beam and an ion beam that have a similar charge density and a drift speed that fulfills the aforementioned condition. 

The Buneman instability does not grow if the background electrons are so hot that their thermal velocity spread exceeds by far their drift velocity relative to the ions. The two-stream instability competes in this case with the ion acoustic instability between both species of the background plasma \cite{Treumann77book}.

If the direction of the wave vector is not constrained to be parallel to the beam velocity vector then the counterstreaming electron beams can drive the predominantly magnetic filamentation instability \cite{Califano98} or the quasi-electric oblique mode instability. Instabilities driven by relativistic electron beams are reviewed by Ref. \cite{Bret10a}.

Beam instabilities have been widely examined in the past both experimentally and theoretically. Many of these studies were performed with particle-in-cell (PIC) simulations. The substantial computational cost of the PIC simulations implied that in some cases the development of the instabilities had to be accelerated by choosing a reduced ion mass. The reduction of the ion mass increases the exponential growth rate of the instabilities in which the ions are involved; it is the Buneman-type instability in the aforementioned case. The ion mass does, however, not affect the instabilities that develop between the counterstreaming electron beams. 

A reduction of the mass of the ions in PIC simulations will thus not only speed up the instability, it will also alter the spectrum of the growing waves. The effects of a reduced ion mass on the exponential growth rate of beam instabilities have been studied systematically in Ref. \cite{Bret10b}. It turns out that in some cases even a moderate reduction of the ion mass can have profound effects on the spectrum of the unstable waves. The process, by which the plasma is thermalized, depends in turn on the instability that saturates first. A reduction of the ion mass can, thus, alter the final state of the plasma with potentially far-reaching consequences. A plasma saturation by the filamentation instability results, for example, in strong magnetic fields \cite{Kazimura98,Honda00,Silva03,Sakai04,Jaroschek05,Nishikawa09,Bret13}, while the other instabilities drive primarily electric fields \cite{Thode73,Dieckmann00}. The Buneman-type instability between the ions and the bulk electrons heats up the latter, while the two-stream instability between the counterstreaming electron beams heats up the beam electrons and possibly the bulk electrons. It is unclear how the saturation  of one instability affects the other. Systematic studies are needed in order to better understand the consequences of using reduced mass ratios not only during the linear growth phase of the instabilities but also after their nonlinear saturation.

Here we test some of the results obtained in Ref. \cite{Bret10b} with PIC simulations, which allow us to explore non-linear effects introduced by the reduced ion mass. We limit ourselves to the mildly relativistic electron speeds, which are representative for solar energetic electrons \cite{Muschietti90,Kontar01,Klein05,Hamish13,Hamish14} and for electrons that have been heated by the ablation of a solid target by a high power laser pulse \cite{Tabak}. 

Numerical artifacts, which are caused by a reduced ion-to-electron mass ratio, become stronger with an increasing relativistic factor of the beam speed \cite{Bret10b}. Some of our results are thus also relevant for numerical studies of interactions between plasma and the more energetic electron beams, which are generated by the wakefield of a laser \cite{Esarey09,Sarri17}. A related study involving ultrarelativistic pair beams can be found in Ref. \cite{Sironi14}. 

Our parametric study is conducted in one spatial dimension and we align the beam velocity vector with the simulation direction, which suppresses the oblique mode instability and the filamentation instability. The results provided by such simulations are realistic if one electron beam is dilute and the second dense and if the beam speeds are not too high \cite{Thode73,Tzoufras06}. 

Our paper is structured as follows. The linear dispersion relation of the plasma is solved and the PIC simulation method is discussed in Section 2.
Section 3 presents the results of our simulation studies and they are discussed in Section 4.

\section{Linear Theory and initial conditions}

\subsection{Linear Theory}

We consider a system composed of a relativistic electron beam with the density $n_{b}$, the reduced velocity $\beta_{b}=v_{b}/c$ and the Lorentz factor 
$\gamma_{b}=\frac{1}{\sqrt{1-\beta_{b}^{2}}}$. The beam crosses a spatially uniform plasma with the densities $n_{i}$ and $n_{e}$ of ions and electrons
and $n_{i}=n_{e}+ n_{b}$. The drift velocity $v_{e}$ of the bulk electrons is such that it cancels out the beam current by $n_{b}v_{b} + n_{e}v_{e}=0$. 
The plasma frequencies of the electrons and ions are $\omega_{pe,pi}=\sqrt{\frac{e^{2}n_{e,i}}{m_{e,i}\epsilon_{0}}}$, where $e$, $m_{e,i}$, 
$\epsilon_{0}$ are the elementary charge, the electron/ion mass and the dielectric constant, respectively. The thermal speed of a species $q$ with the mass $m_q$ 
and temperature $T_q$ is $v_{Tq}={(k_BT_q/m_q)}^{1/2}$ ($k_B:$ Boltzmann constant). Time is normalized by $\omega_{pe}^{-1}$, space by $c\omega_{pe}^{-1}$ 
and frequencies $\omega$ by $\omega_{pe}$. 

For the stability analysis, we consider the response of the system to harmonic perturbations 
$\varpropto exp(i\textbf{k}.\textbf{r}-\omega t)$. We reduce the system to one spatial dimension (x direction), we align the simulation direction with the beam drift velocity 
and define the normalized variables
\begin{equation}
R=\frac{m_{i}}{m_{e}}, \quad Z=\frac{k_{x} v_{b}}{\omega_{pe}} ,\quad  \alpha=\frac{n_{b}}{n_{e}}, \quad \Omega = \frac{\omega}{\omega_{pe}}.
\end{equation}
We assume that the thermal speeds of both electron species are small compared to $v_b$ and that the thermal speeds of the bulk ions and electrons are small 
compared to $v_e$. The dispersion equation $K_{L}(Z,\Omega)$ for this cold plasma is given, for example, in Ref. \cite{Bret10b}. Its Eigenmodes fulfill the 
dispersion relation $K_{L}(Z,\Omega)=0$ or 
\begin{equation}
1- \underbrace{\frac{1+\alpha}{R\Omega^2}}_{\text{Ion current}}- \underbrace{\frac{\alpha}{\gamma_{b}^{3}(\Omega -Z)^2}}_{\text{beam current}} - \underbrace{\frac{1}{\gamma_{e}^{3}(\Omega +\alpha Z)^2}}_{\text{return current}} = 0.
\label{dispe_Bret}
\end{equation}

Figure \ref{TSI_Bret} shows the exponential growth rate $ \delta \equiv Im(\Omega)$ of the instability, which is obtained  from the numerical solution of Eqn. \eqref{dispe_Bret}, as a function of $R$ for $\gamma_{b}=2$ and for the values $\alpha=0.3, 0.03$ and $0.003$. 
\begin{figure*}[htb]
 \subfloat[]{\includegraphics[width=0.34\textwidth]{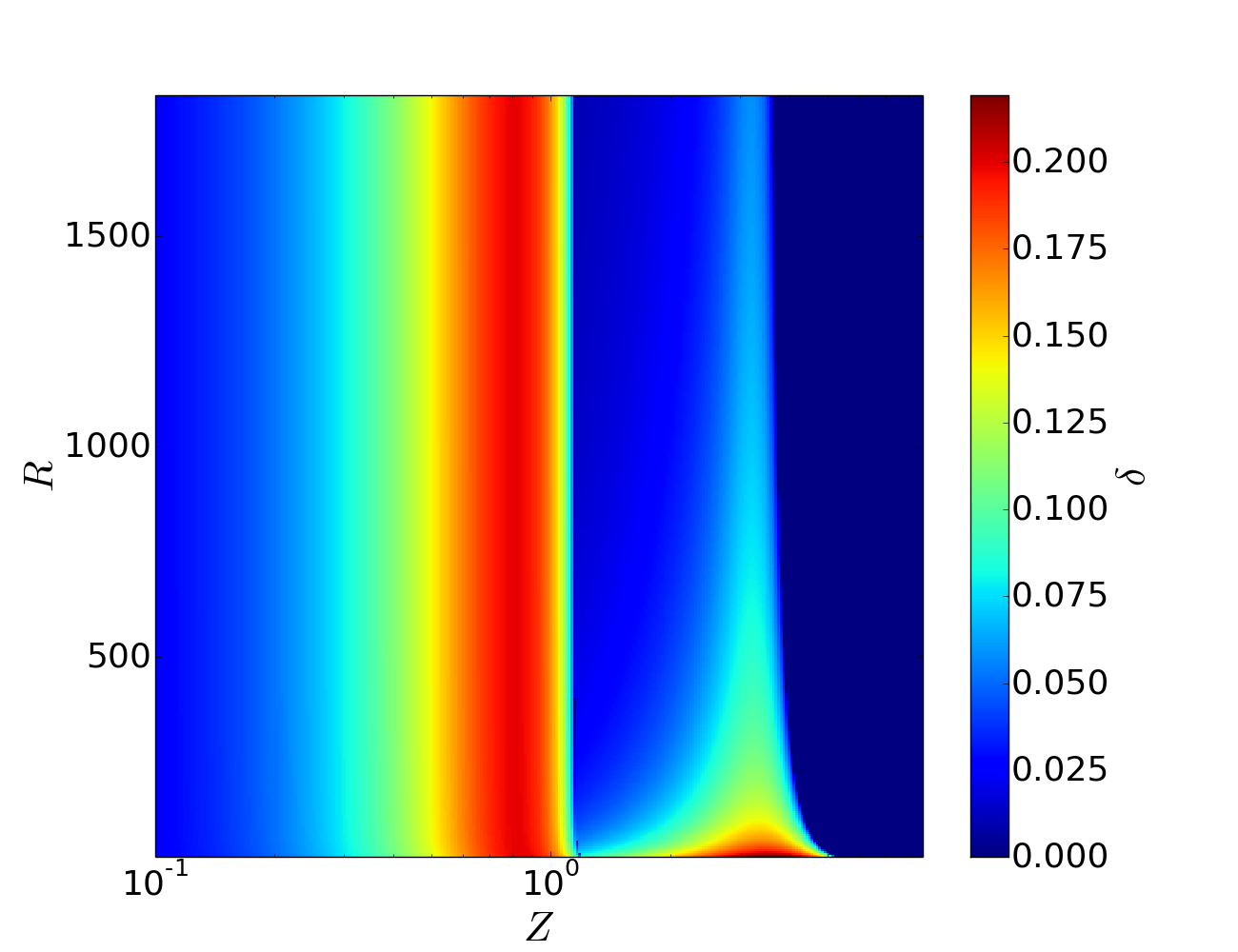}\label{im1}}
 \subfloat[]{\includegraphics[width=0.34\textwidth]{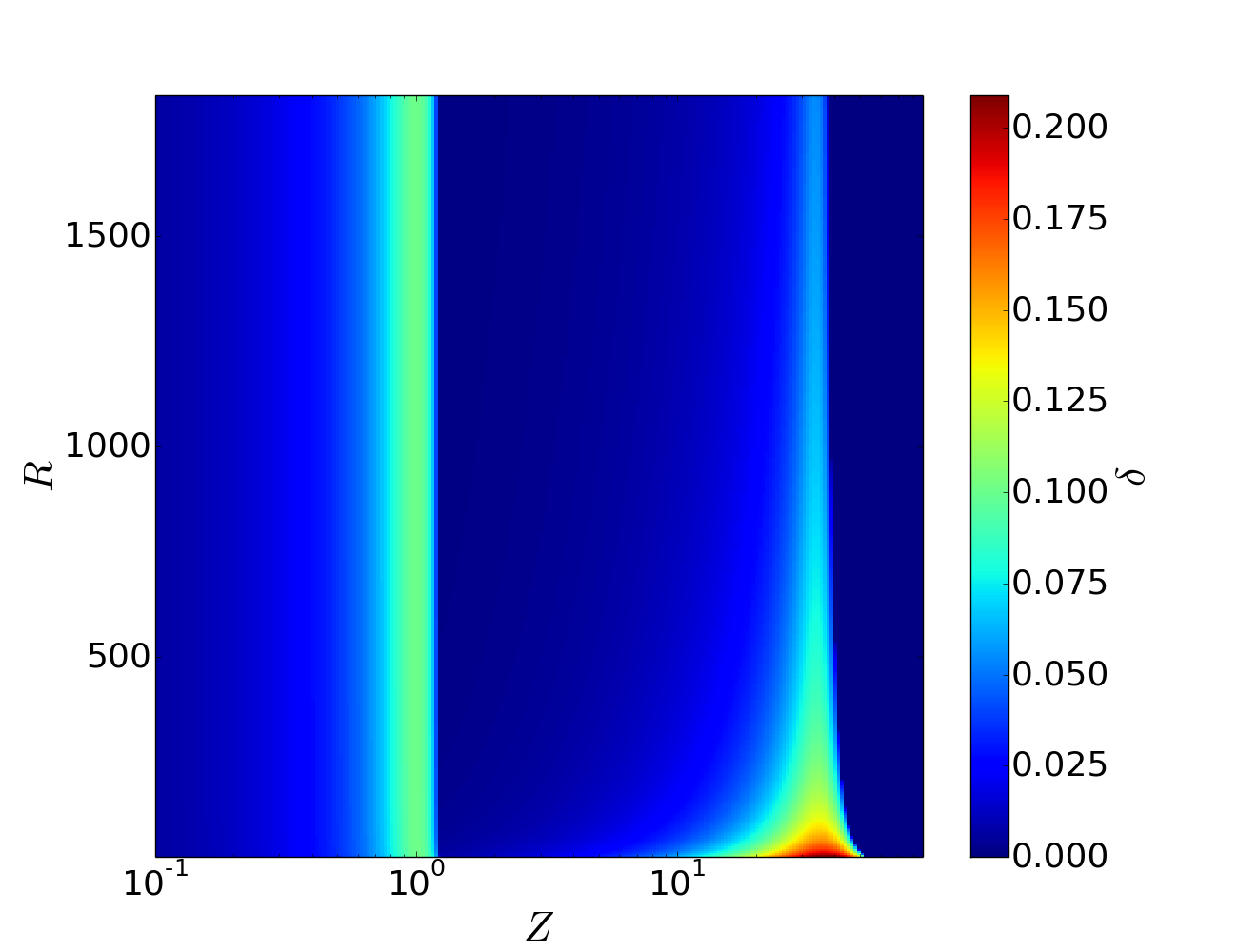}\label{im2}}
 \subfloat[]{\includegraphics[width=0.34\textwidth]{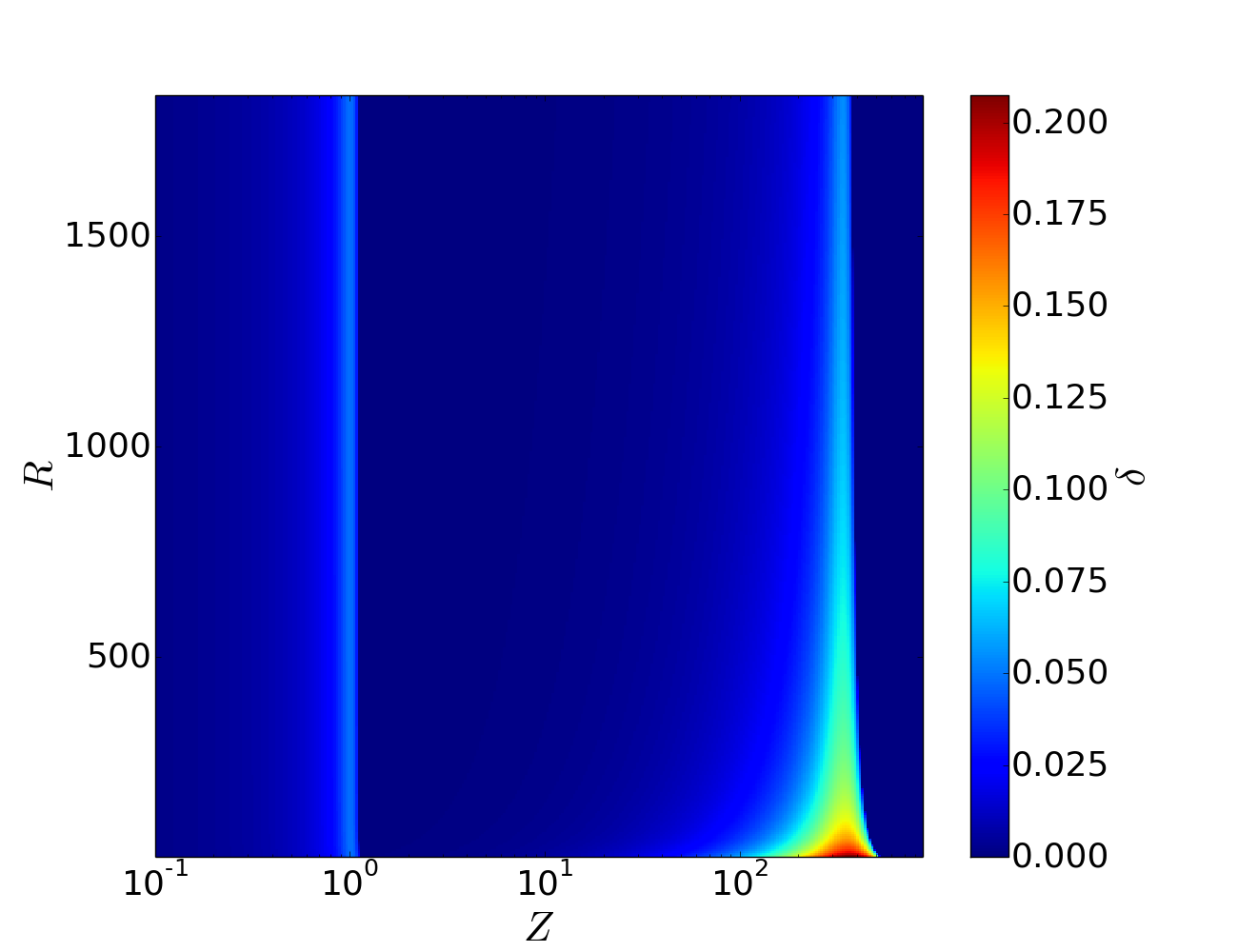}\label{im3}} 
 \caption{Exponential growth rates $\delta (Z,R)$ obtained from the solution of the linear dispersion relation Eqn. \eqref{dispe_Bret} for $\gamma_{b}=2$. 
 The growth rates for $\alpha=0.3, 0.03$ and $0.003$ are shown in panel (a), panel (b), and panel (c) respectively. \label{TSI_Bret}}
\end{figure*}
This instability is the superposition of the electron two-stream instability and the Buneman instability. The first one at low $Z$ is driven by the electrons 
of the beam and the electronic return current. The second arises from the interaction of the electronic return current with the ions. In the limit 
$\alpha<<1$ the two-stream instability has its maximum growth rate $\delta $ at the wavenumber $Z \sim 1$, with $\delta \sim \frac{\sqrt{3}}{2^{4/3}} 
\frac{\alpha^{1/3}}{\gamma_{b}}$, and the unstable Buneman instability at the wavenumber $Z \sim 1/ \alpha$, with $\delta \sim \frac{\sqrt{3}}{2^{4/3}} R^{-1/3}$. 

For dense beams, the growth rate of the two-stream instability is largest regardless of $R$. However, when $\alpha$ and $R$ are low, the growth rate of the Buneman instability exceeds that of the two-stream instability. Table \ref{thetable} lists the maximum of the growth rate $\delta$, localized at $Z \sim 330$, for $\alpha=0.003$ and for three different values of $R$. Figure \label{TSI_Bret} demonstrates that the choice of $R$ affects most strongly the beam with $\alpha = 0.003$ and we focus on this case.

\begin{table}
\begin{tabular}{|*{5}{c|}l r|}

   & ion mass $R$ (in $m_e$) & 1836 & 400 & 25  \\
  \hline
  $\alpha = 0.3 $ & maximum of $\delta$ & 0.20 & 0.20 & 0.22  \\
  \hline
  $\alpha = 0.03 $ & maximum of $\delta$ & 0.10 & 0.10 & 0.21  \\
  \hline
  $\alpha = 0.003 $ & maximum of $\delta$ & 0.058 & 0.089 & 0.207  \\
  
\end{tabular}
\caption{The growth rate of the Buneman instability for three different values of $R$ and for $\alpha$.} 
\label{thetable}
\end{table}

\subsection{The PIC code and the initial conditions}

We use the particle-in-cell (PIC) simulation code EPOCH \cite{EPOCH}. It solves the Vlasov-Maxwell system of equations via the method of characteristics. 
Amp\`ere\textquotesingle s and Faraday\textquotesingle s law are solved on a grid and the code fulfills Gauss's law and $\nabla \cdot \textbf{B} =0$ to round-off precision. 
The plasma is approximated by an ensemble of computational particles (CPs). The momentum of each CP is updated via a discretized form of the Lorentz 
force equation, which uses the electromagnetic field values that have been interpolated from the numerical grid to the position of the CP. The current of each 
CP is deposited on the numerical grid using Esirkepov's scheme \cite{Esirkepov00CPC}. 
We resolve one spatial dimension and the three velocity components of the CP’s (1D3V).

The simulation domain is resolved by $N_{x}=9\times 10^4$ cells. The density of the dilute electron beam is $n_{b}=0.003 \: n_{i}$ and its Lorentz factor  
$\gamma_{b}=2$. The density of the background electrons is $n_{e}=0.997 n_{i}$ and their mean velocity $\beta_{e}=0.0026$. The length of the box is 
$L_{x} = 31.4$ and its spatial resolution is $\Delta x=3.5 \times 10^{-4}$. Periodic boundary conditions for the electromagnetic fields and for the CPs are 
used. The maximum resolved wave number $Z^{max}$ and its resolution $\Delta Z$ are respectively $Z^{max}= \beta_e \frac{\pi}{\Delta x} =7800$ and 
$\Delta Z=\beta_e \frac{2 \pi}{L_{x}}= 0.17$. The electrons and the ions have Maxwellian velocity distributions. We vary the ion mass and the temperatures 
of all species in a range where all beams are practically cold, so to respect the assumptions underlying Eq. \ref{dispe_Bret}. The temperature of the electron beam 
is $T_{b}= 10 eV$, which gives the thermal speed $v_{Tb} \approx  5 \times 10^{-3} v_{b}$. The temperature of the bulk electrons 
is $T_{e}= 0.1 eV$, which gives $v_{Te} \approx 0.2 v_{e}$. The Debye length $\lambda_{De}= v_{Te}\omega_{pe}^{-1}$ is $\lambda_{De}= 1.25 \Delta x$. We represent the ions by 250 
particles per cell (ppc), the bulk electrons by 200 ppc and the beam electrons by 50 ppc. 

\section{Simulation results}

\subsection{The linear wave growth and its saturation}

\begin{figure*}[htb]
 \subfloat[]{\includegraphics[width=0.34\textwidth]{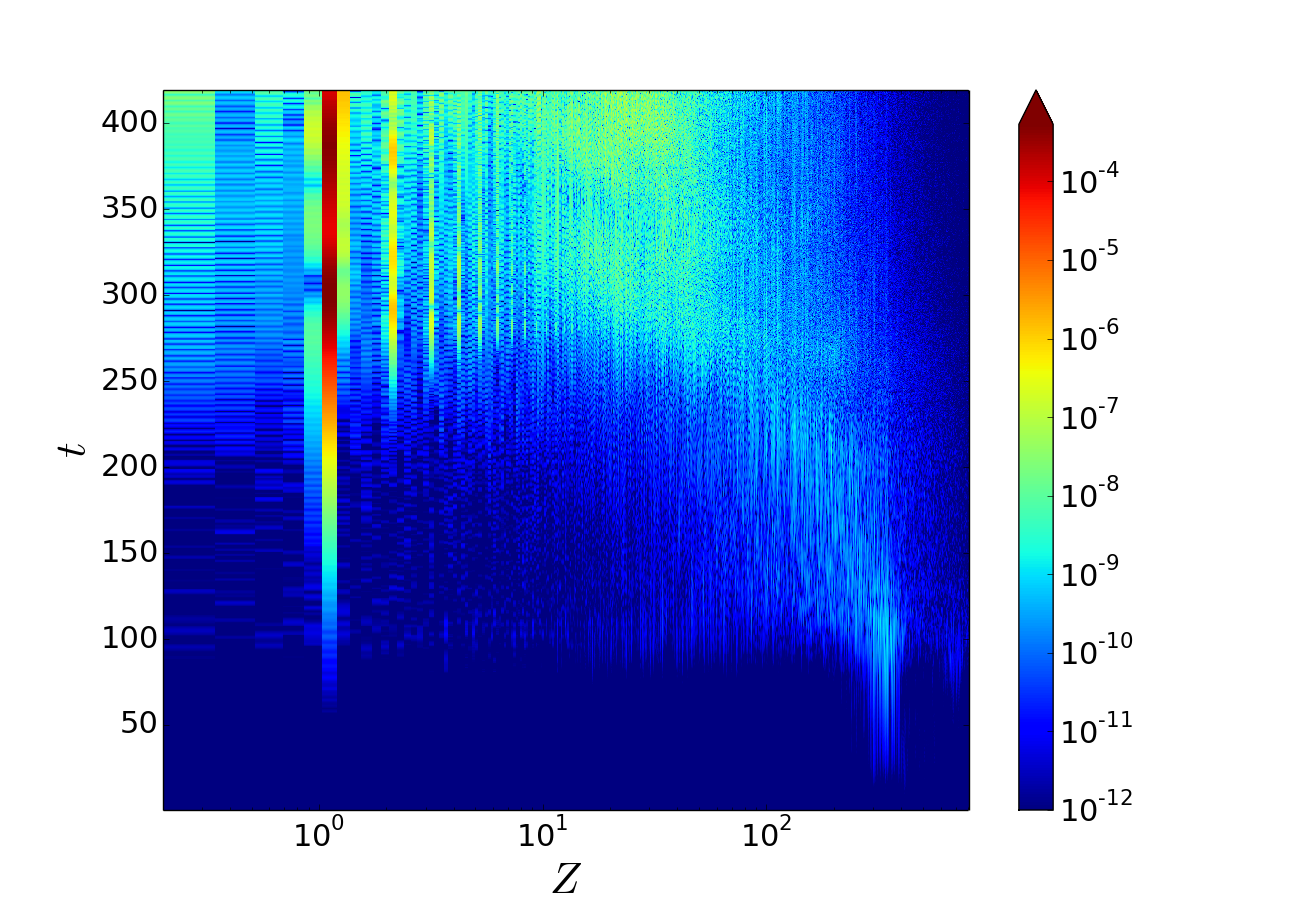}\label{tkx_mi1836}}
 \subfloat[]{\includegraphics[width=0.34\textwidth]{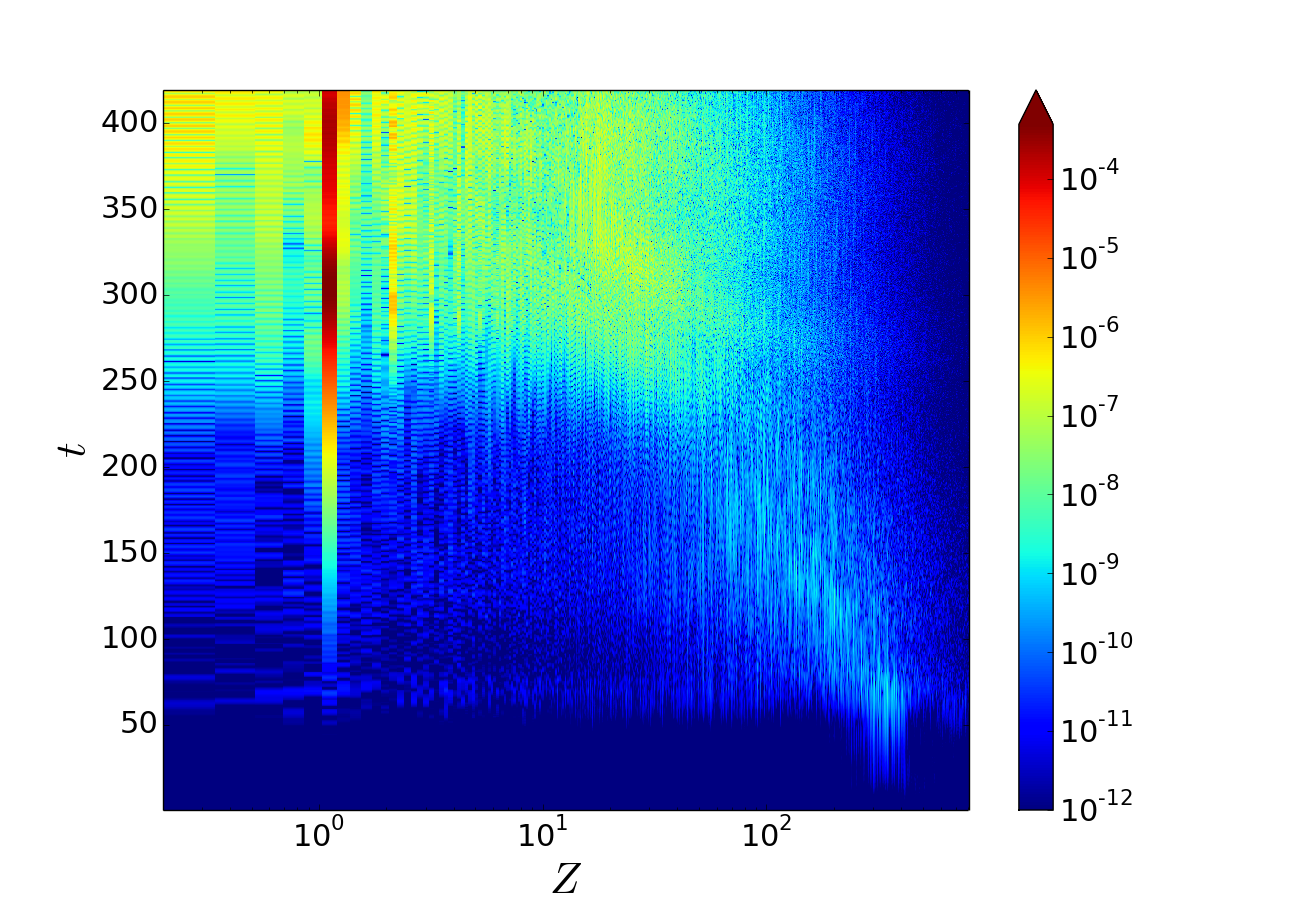}\label{tkx_mi400}}
 \subfloat[]{\includegraphics[width=0.34\textwidth]{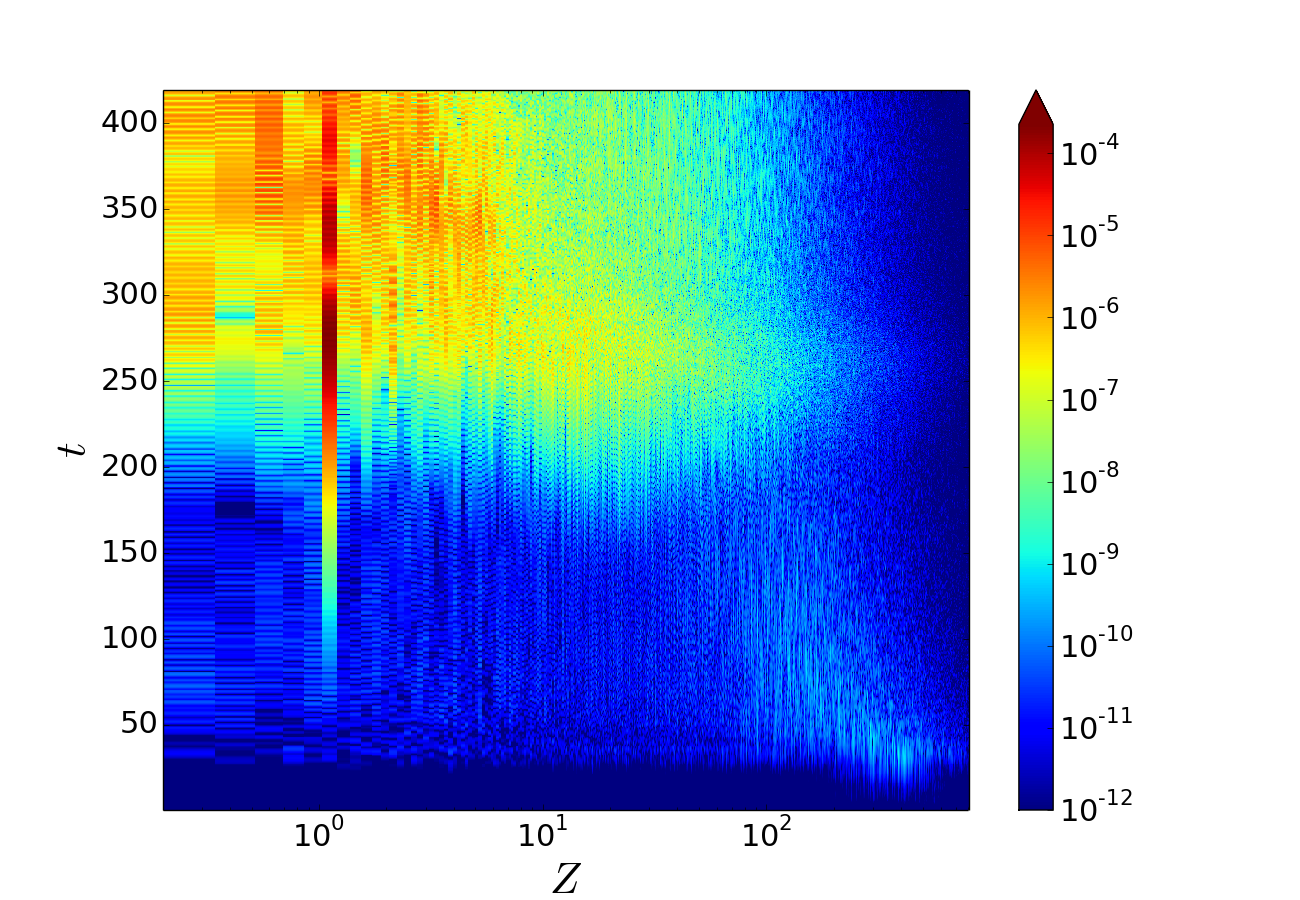}\label{tkx_mi25}}
\caption{The power spectrum $P_x(Z,t)$ of the electric field component $E_x$ normalized to $m_e^2c^2\omega_{pe}^2/e^2$: panel (a) corresponds to $R=1836$, panel (b) to $R=400$ and (c) to $R=25$. The color scale is 10-logarithmic. \label{t_kx_2ga5_mi1836_T1} }
\end{figure*}

We compare the range of unstable wave numbers obtained from the solution of Eqn. \ref{dispe_Bret} (See Fig. \ref{TSI_Bret}) with that of the waves in the PIC simulations and determine the saturation time as a function of $R$. 
We analyze the electric field component $E_{x}(x, t)$, which grows in response to the two-stream instability and Buneman instability, by performing a Fourier transform over space
\begin{equation}
E_{x}(j\Delta Z,t)= N_{x}^{-1} \sum \limits_{p=1}^{N_{x}} E_{x}(p\Delta x,t) e^{-j p \Delta x \Delta Z}.
\end{equation}
Figure \ref{t_kx_2ga5_mi1836_T1} shows the power spectra $P_{x}(Z,t)={|E_x(Z,t)|}^2$ for the values $R=$ 1836, 400 and 25. 
The wave numbers $ Z_{Bun} \sim 333 $ and $Z_{TSI} \sim 1$, where the growth rates of the Buneman instability and the two-stream instability reach their maxima according to Eqn. \ref{dispe_Bret}, coincide with the values where the instabilities grow in Fig. \ref{t_kx_2ga5_mi1836_T1}. 
The instabilities start to grow after about $t\approx 15$.
This delay can be attributed to the time required by the thermal noise that seeds the instabilities to grow and to the need to establish a coherent wave along the beam direction. 

Figure \ref{t_kx_2ga5_mi1836_T1} shows that both instabilities grow and saturate independently. For this low value of $\alpha$ the growth rate of the Buneman instability is comparable to that of the two-stream instability if $R=1836$ and larger for smaller $R$. Indeed the Buneman mode at $Z \approx Z_{Bun}$ reaches about the same power as the two-stream mode at $Z \approx Z_{TSI}$ at $t\approx 100$ in Fig. \ref{t_kx_2ga5_mi1836_T1}(a) while it outgrows the two-stream modes in the cases $R=400$ and $R=25$. 

The field power at $Z \approx Z_{TSI}$ evolves similarly in all three simulations on the displayed time interval; the evolution of this instability is unaffected by the value of $R$ during its linear growth phase. We observe several harmonics of the wave at $Z \approx Z_{TSI}$ for $R=1836$ and one for $R=400$. Only a broad wave continuum is observed for $R=25$. The peak power of the low-Z mode and the number of observed harmonics increases with $R$, which shows that the wave can sustain a sine shape for a larger amplitude and for a longer time. 

The waves driven by the Buneman instability are not stable. Once the wave power at $Z_{Bun}$ has peaked the interval, in which the wave power is concentrated, is shifted in time to lower values of $Z$. The waves driven by the Buneman instability are amplified after $t \approx 250$ for all $R$ by their coupling to the two-stream mode. This coupling is responsible for the onset of the broadband electrostatic wave activity, which is particularly strong for the cases $R=400$ and $R=25$. 

\begin{figure}[!h]
	\begin{center}
		\includegraphics[scale=0.24]{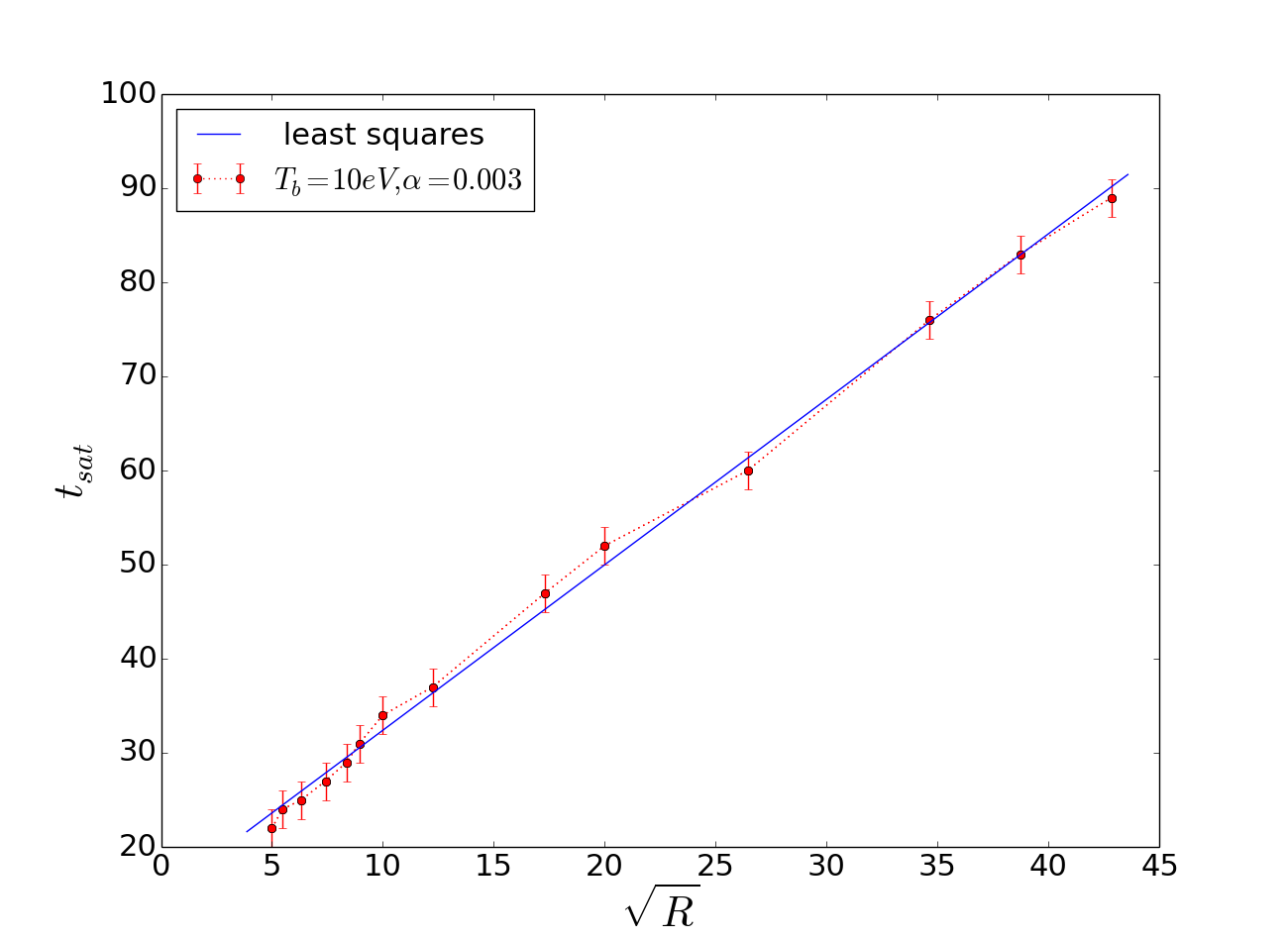}
		\caption{Scaling of the saturation time with $R$. The temperature of the electron beam is $T_{b}= 10 eV$ and the temperature of the bulk electrons is $T_{e}= 0.1 eV$. \label{scaling_tsat}}
	\end{center}
\end{figure}

In what follows we test if the dependence of the exponential growth rate of the Buneman instability on $R$ is the only reason for its faster saturation with decreasing $R$. The exponential growth rate of the Buneman instability is $\delta \sim \frac{\sqrt{3}}{2^{4/3}} R^{-1/3}$. According to this growth rate the wave amplitude will reach a given amplitude after a time $t\propto R^{1/3}$, provided that the seed electric field for the instability does not depend on $R$. We have performed a parametric study of the saturation time of the Buneman instability as a function of $R$ in order to determine its scaling with the ion mass. The saturation time of the Buneman instability fulfills $t_{sat} -15 \propto  \sqrt{R}$ as shown in Figure \ref{scaling_tsat}. The subtracted time 15 corresponds approximately to the delay of the wave growth observed in Fig. \ref{t_kx_2ga5_mi1836_T1}. The scaling of the saturation time $\propto \sqrt{R}$ does not match the scaling $t\propto R^{1/3}$ of the time it takes the Buneman wave to reach a given amplitude. The Buneman instability reaches its nonlinear regime much faster for a low value of $R$ than for a large one.

\subsection{Non-linear saturation and energy transfer}

We explore in this section how the value for $R$ affects the energy exchange between the three plasma species and the electric field. 
We integrate for this purpose the energy density of the electric field's $E_x$ component and the energy densities of the individual particle species over the entire simulation box. 
Particle energies are measured in the reference frame, in which the total momentum vanishes at $t=0$, which coincides with the simulation frame. 
We normalize all energies to the total energy.

\begin{figure*}[htb]
 \subfloat[]{\includegraphics[width=0.34\textwidth]{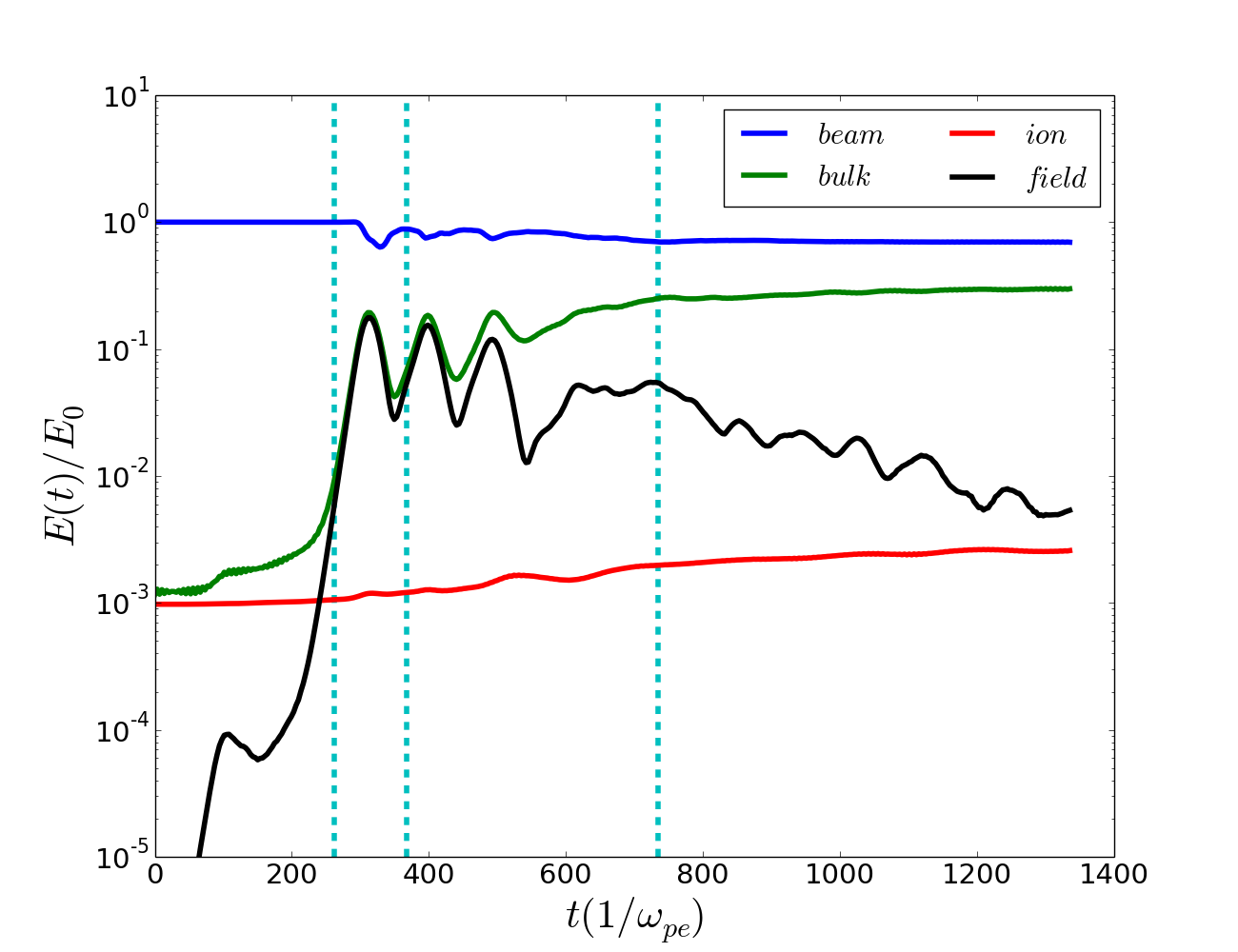}}
 \subfloat[]{\includegraphics[width=0.34\textwidth]{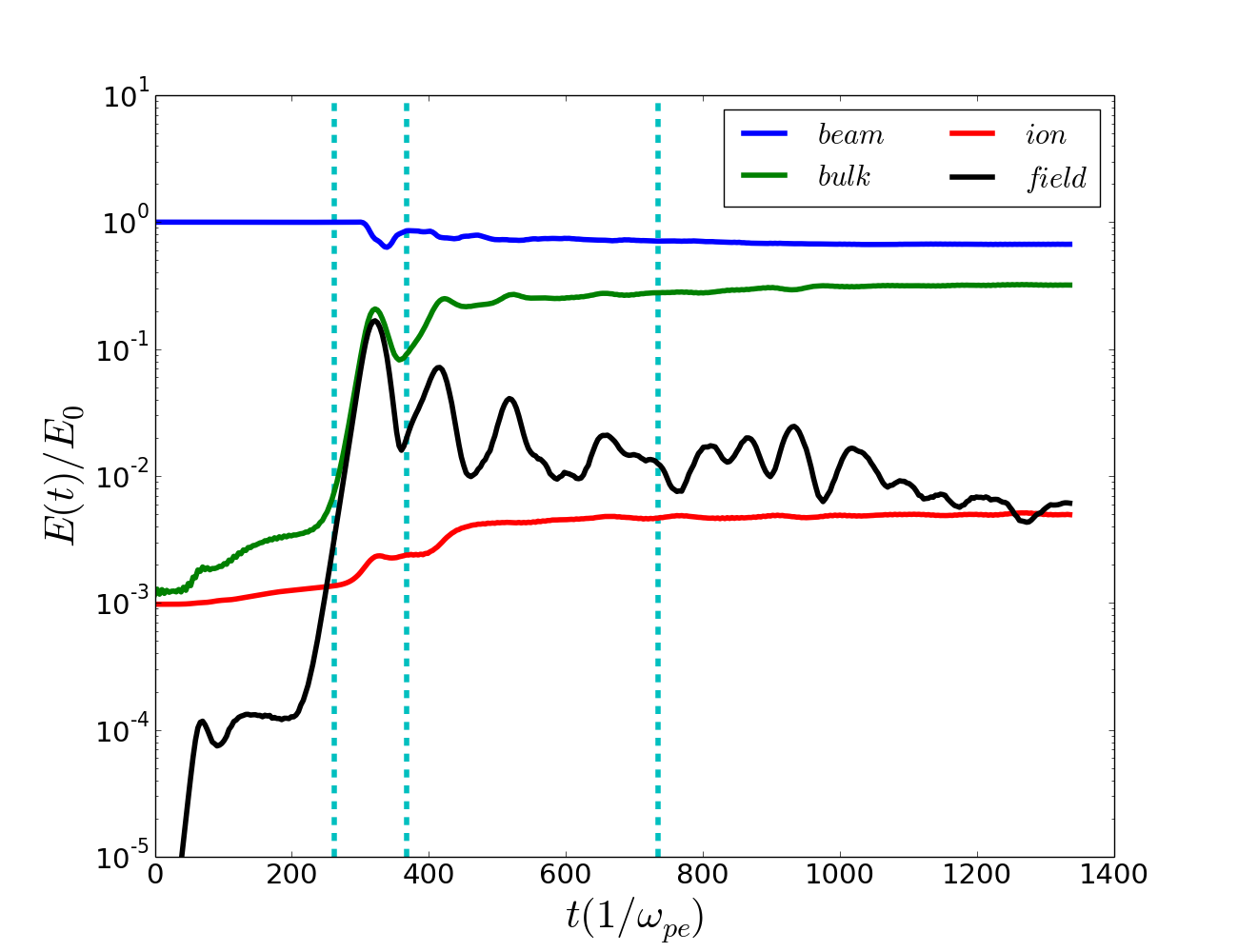}}
 \subfloat[]{\includegraphics[width=0.34\textwidth]{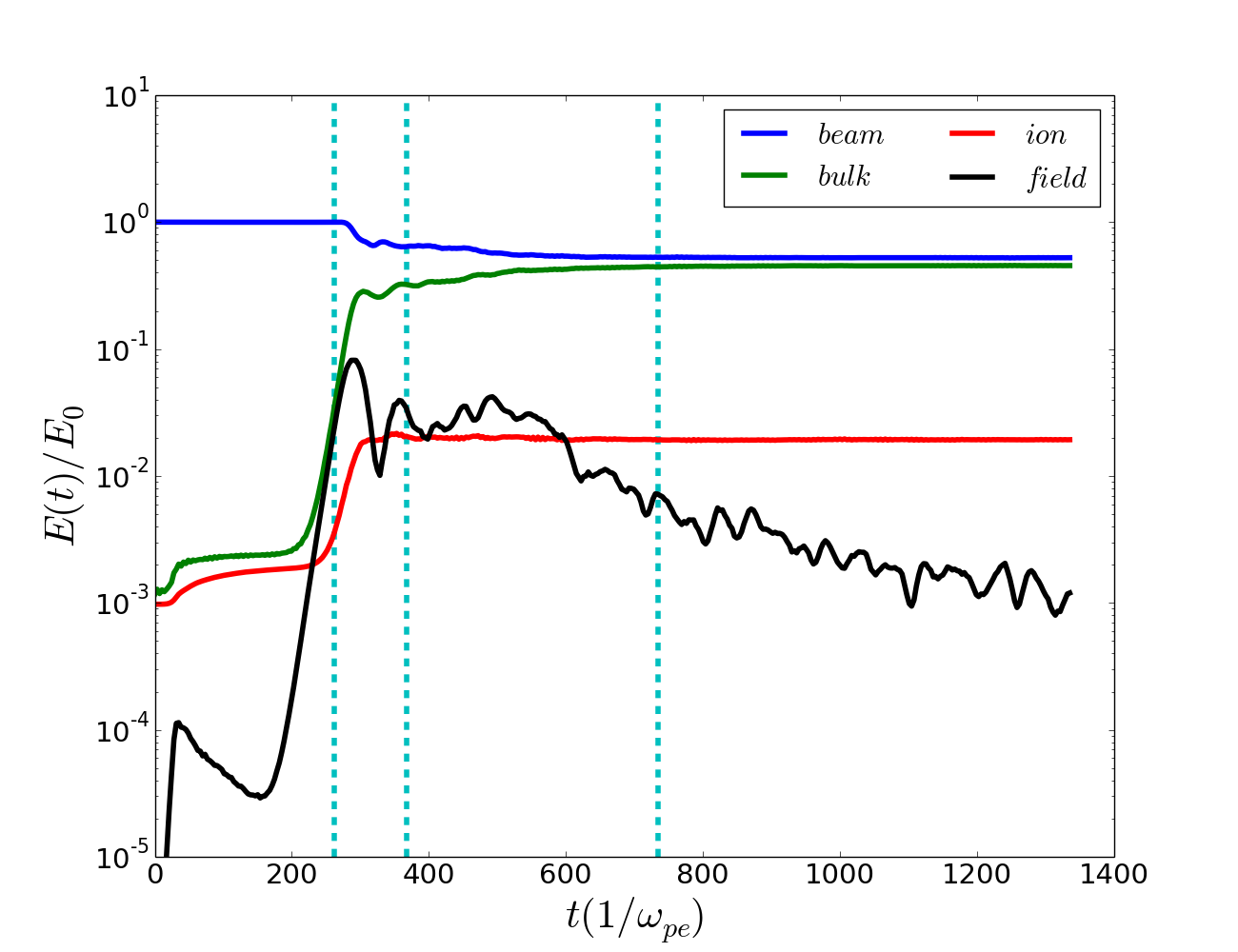}}
 \caption{Panel (a) shows the particle and field energies for a mass ratio $R =1836 $, panel (b) for a mass ratio $400$ and panel (c) for the mass ratio $25$. 
 The temperature of the electron beam is $T_{b}= 10 eV$ and the temperature of the bulk electrons is $T_{e}= 0.1 eV$, the dashed lines correspond to the times $t=$ 262, 367 and 734.}\label{density_ener}
\end{figure*}

Figure \ref{density_ener} shows the time evolution of all energies for the three mass ratios 1836, 400 and 25.
The electric field energy grows exponentially at early times and saturates at $t\approx 90$ in Fig. \ref{density_ener}(a), at $t=50$ in Fig. \ref{density_ener}(b) and at $t\approx 25$ in Fig. \ref{density_ener}(c). The faster rise of the field energy at low $R$ reflects the larger growth rate of the Buneman instability. The field energies at the time the Buneman instability saturates are the same in all simulations, which suggests that the saturation is caused by the interaction between the wave and electrons. This in confirmed by Fig. \ref{density_ener}. The energy the ions gain when the Buneman instability saturates does increase with decreasing $R$ but it remains small compared to the energy gain of the bulk electrons. According to Fig. \ref{scaling_tsat} the Buneman instability saturates earlier than expected for low $R$, which suggests an involvement of the ions in the saturation, while Fig. \ref{density_ener} demonstrates that the instability saturates by its interaction with the electrons. Both observations are compatible if $R$ affects the phase speed of the electrostatic wave relative to the electrons. A decrease of the phase speed in the rest frame of the electrons with $R$  implies that the latter can react more easily to the wave.

The energies of the electric field, the ions and the bulk electrons all increase in Fig. \ref{density_ener} when the Buneman instability saturates. Energy conservation implies that the saturation of the Buneman instability must have extracted energy from the beam electrons. The electric field energy decreases after the saturation of the Buneman instability and this energy decrease depends on $R$. 

The electric field energy rises again after $t=200$. Its growth rate is the same in all three simulations as expected for the two-stream instability. 
The electric field energy saturates at $t\approx 300$ and reaches a peak value that is about 20\% of the beam electron energy for $R=$ 1836 and 400. 
The two-stream instability saturates earlier and at a lower peak value for $R=25$. 
The phase speed of the wave, which is driven by the two-stream instability, does not depend on $R$. This wave will enforce a stronger reaction of ions with a low value of $R$. 
We have observed this stronger reaction already in Fig. \ref{t_kx_2ga5_mi1836_T1}(c), which revealed a lower peak power of the electric field and strong  broadband that set in when the wave saturated. 
Figure \ref{density_ener} also shows that the bulk electron energy closely follows that of the electric field until its saturation, after which the individual energies evolve differently in all simulations. 

The energies of the electric field and of the bulk electrons oscillate in phase three times for $R=1836$. The beam electron energy oscillates in antiphase. 
The ion energy hardly reacts to these oscillations for $R=$ 1836. 
The ions with $R=400$ are accelerated by the electric field, which damps the energy oscillations of the electrons and the electric field.
A reduction of the ion mass to $R=25$ boosts their response to the electric field and the damping of the oscillations of the energies of the electrons and the electric field. 
The energies of the bulk electrons and of the beam electrons converge in all simulations and they become almost equal for $R=25$. 
However, the mean energy per electron is still much larger for the beam particles since $\alpha \ll 1$. 
At the simulations's end, the ion energy in the simulation with $R=25$ exceeds that in the simulation with $R=1836$ by an order of magnitude. 

\begin{figure}[!h]
	\begin{center}
		\includegraphics[scale=0.24]{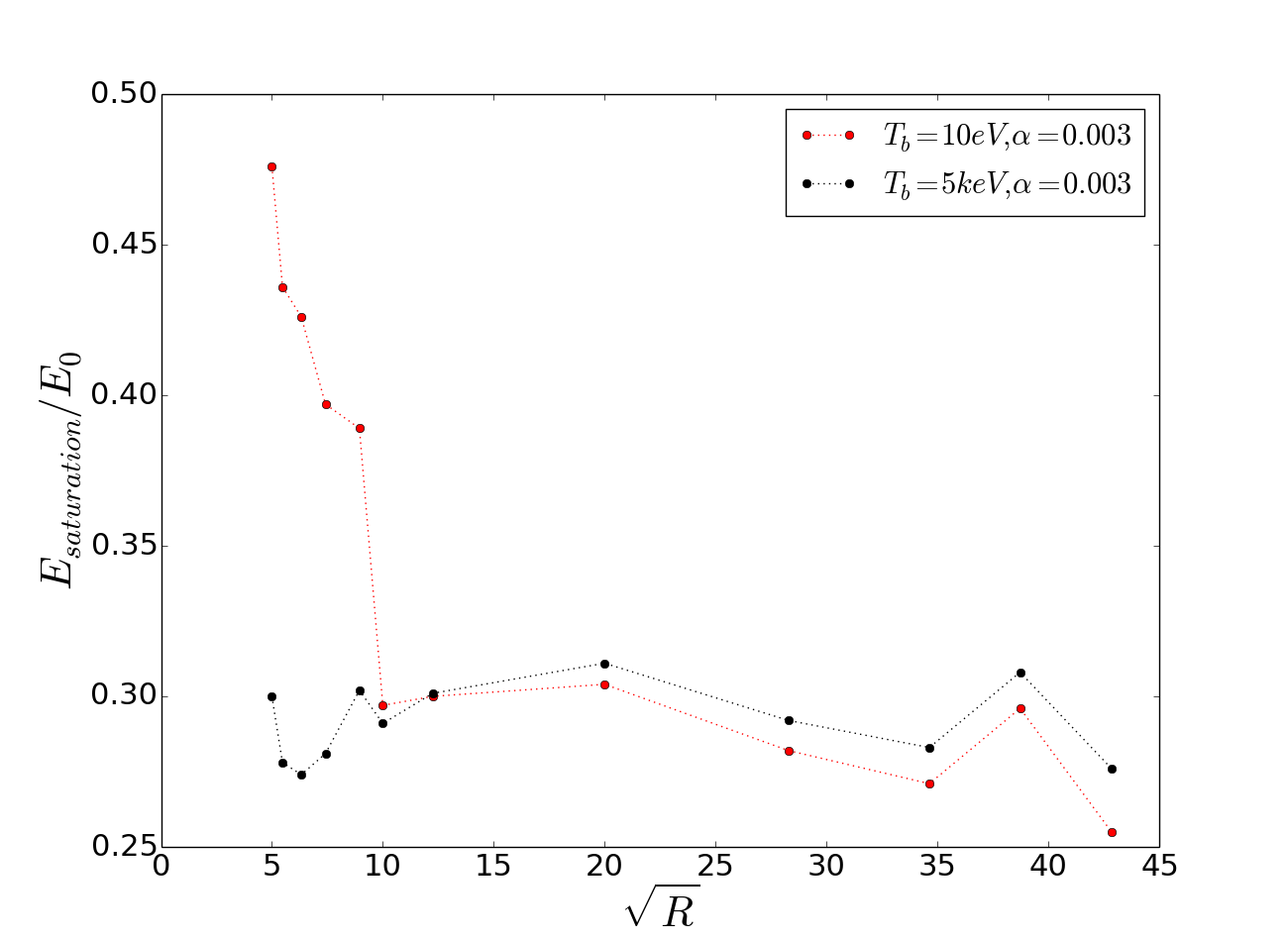}
		\caption{The saturation energy of the bulk electrons at $t= 1250$ in units of the total energy for $\alpha=  0.003$ and for the temperature $T_{b}= 10 \: eV$ for the electron beam. The temperature of the bulk electrons is $T_{e}= 0.1 eV$ \label{Esat_mi_a0003_sqrt_paper} }
	\end{center}
\end{figure}

The simulations, which provided the data shown in Fig. \ref{scaling_tsat}, were followed over a longer time in order to determine the efficiency of the energy transfer from the electron beam to the bulk electrons as a function of $R$. 
Figure \ref{Esat_mi_a0003_sqrt_paper} shows their results.
The bulk electrons gain between 25\% and 30\% of the total energy after the two-stream instability's saturation if $R>100$. 
The energy increases from 30\% to almost 50\% for decreasing values $R<100$. A reduction of $R$ thus enhances the transfer of energy from the electron beam to the bulk electrons.  
This effect is, however, only observed if the electron beam is cold. Increasing the temperature of the electron beam from 10 eV to 5 keV suppresses the rise of the final energy of the bulk electrons at low values of $R$. 

\subsection{Phase space density distributions}

\begin{figure*}[htbp]
\hspace*{-1.5cm}
\begin{tabular}{*{4}{c}}{\centering}
    & \multicolumn{3}{c}{$m_{i}=1836 \: m_{e}$} \\
    \textcolor{blue}{beam} & \subfloat[]{\includegraphics[width=0.34\textwidth]{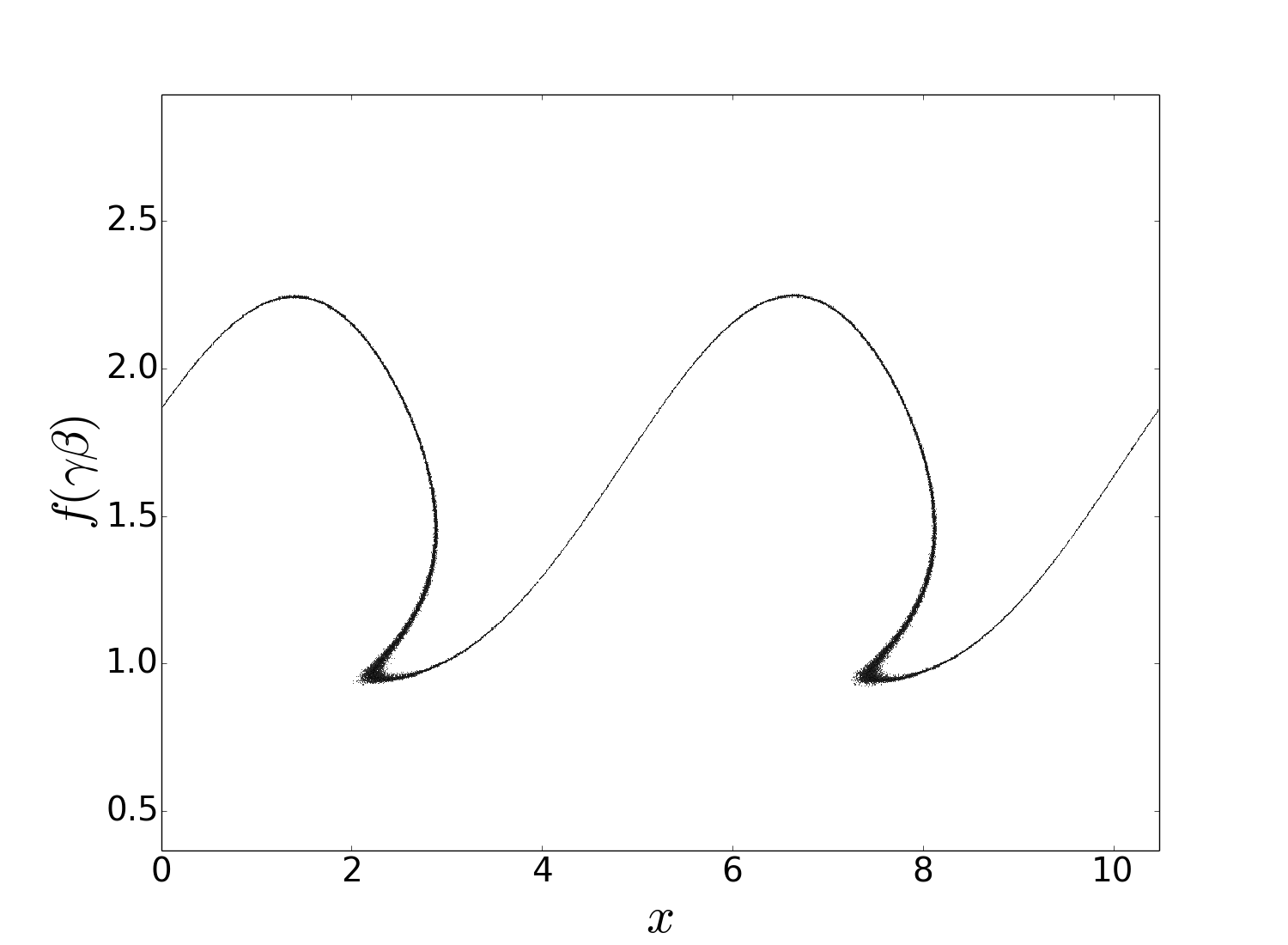}
    \label{2ga_mi1836_Tb10eV_Te01eV_xpx_262wpe_beam}} 
    & \subfloat[]{\includegraphics[width=0.34\textwidth]{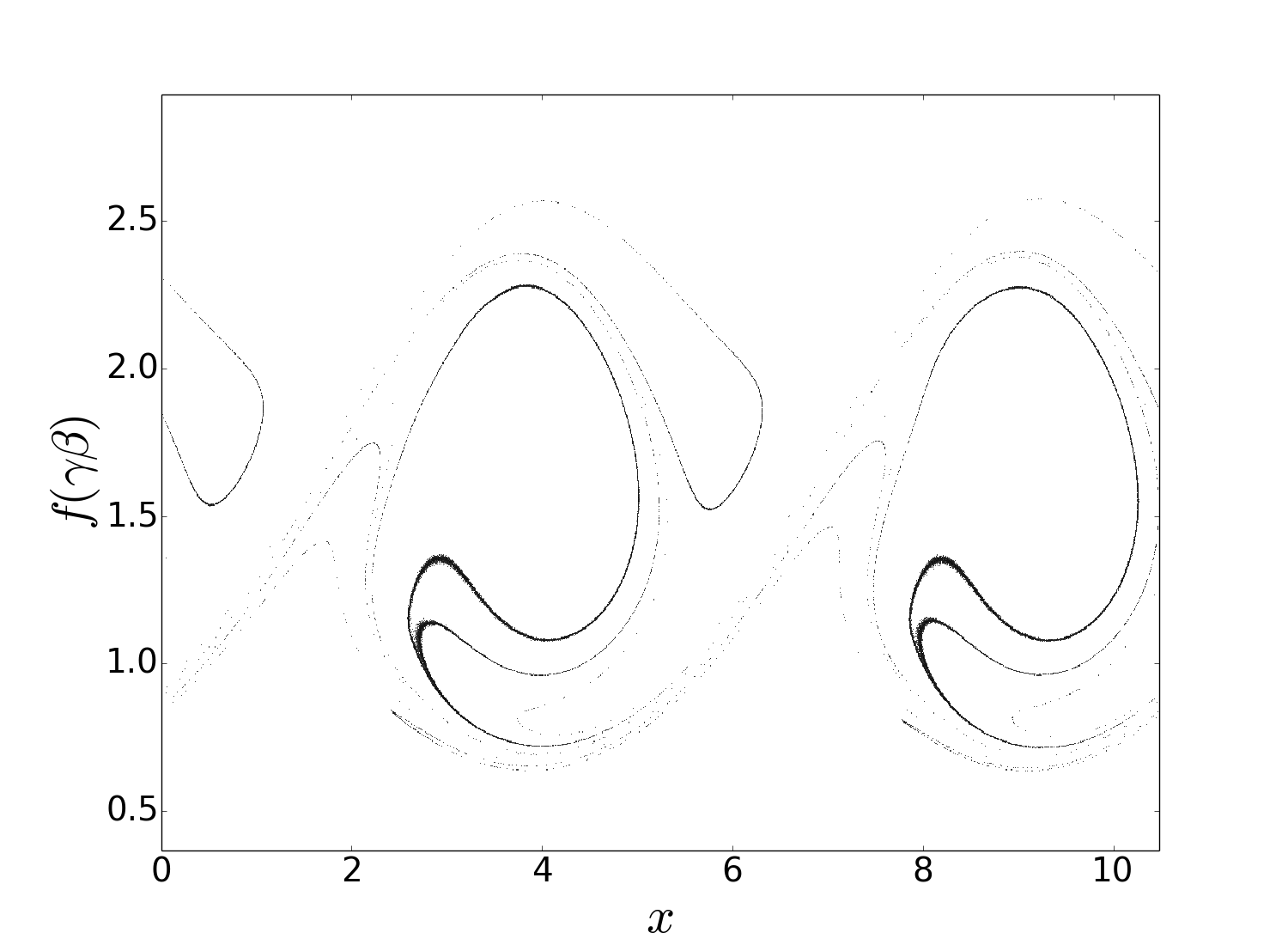}
    \label{2ga_mi1836_Tb10eV_Te01eV_xpx_367wpe_beam}} 
    & \subfloat[]{\includegraphics[width=0.34\textwidth]{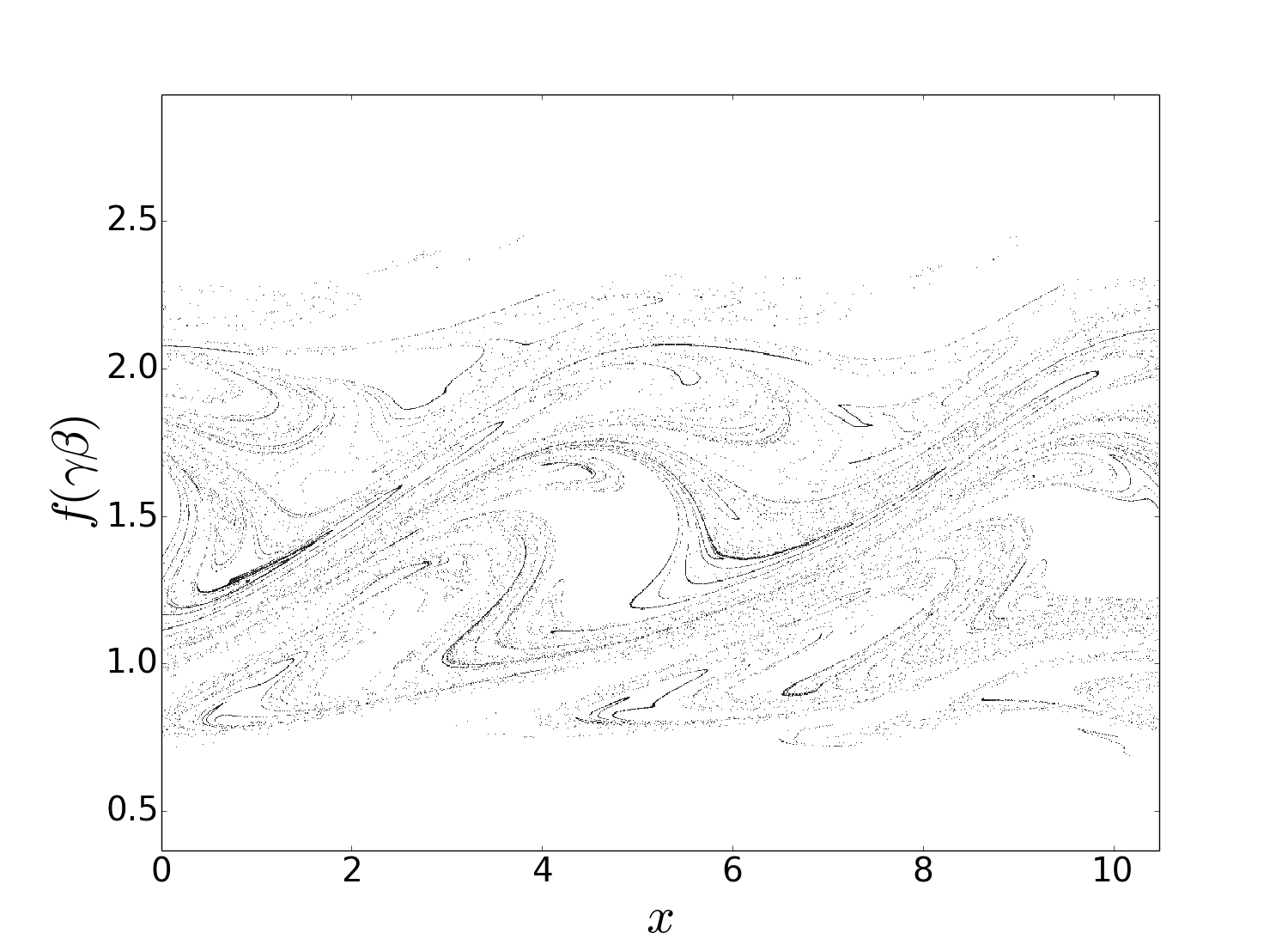}
    \label{2ga_mi1836_Tb10eV_Te01eV_xpx_734wpe_beam}} \\
    \textcolor{green}{bulk} & \subfloat[]{\includegraphics[width=0.34\textwidth]{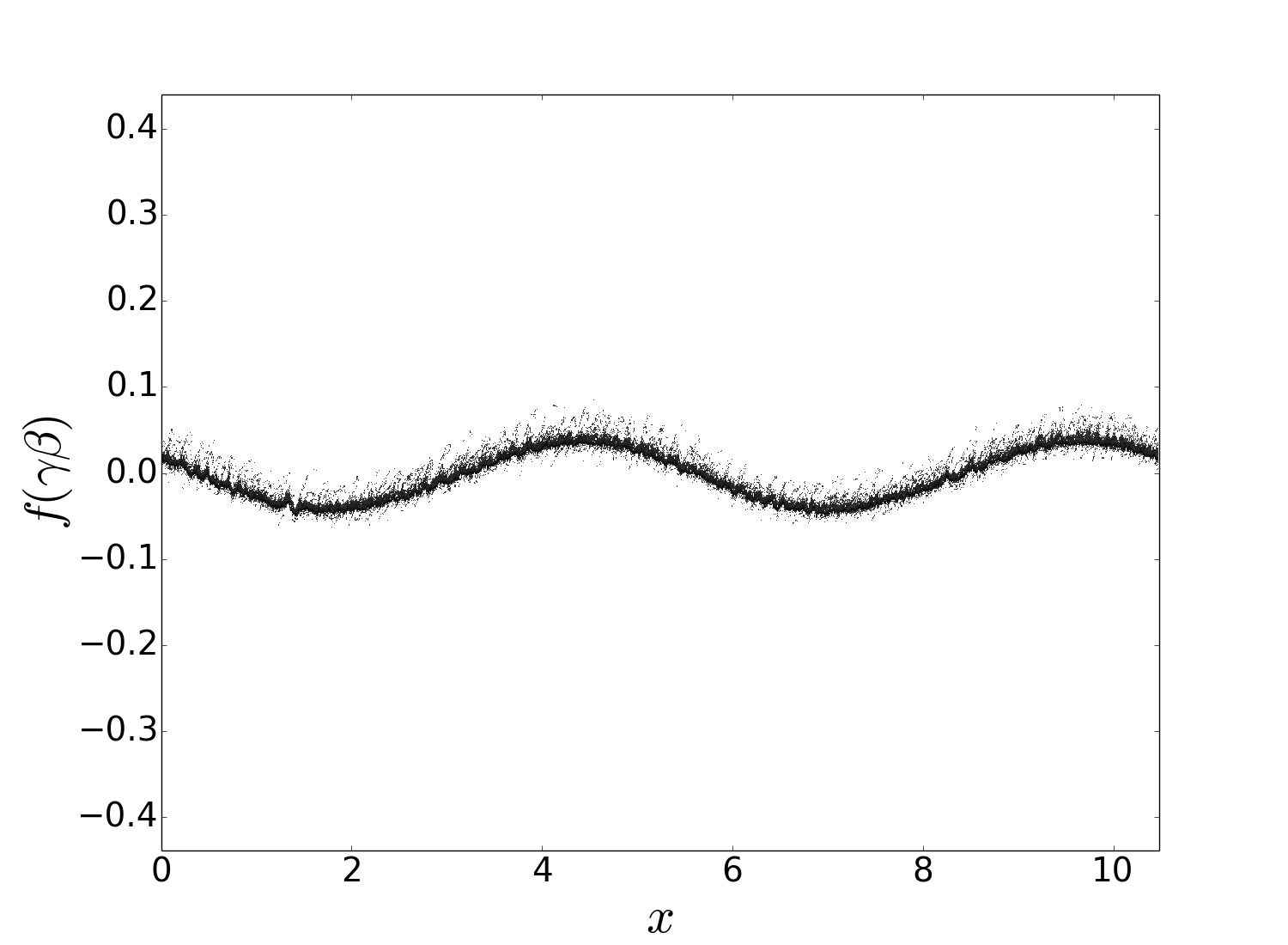}
    \label{2ga_mi1836_Tb10eV_Te01eV_xpx_262wpe_bulk}} 
    & \subfloat[]{\includegraphics[width=0.34\textwidth]{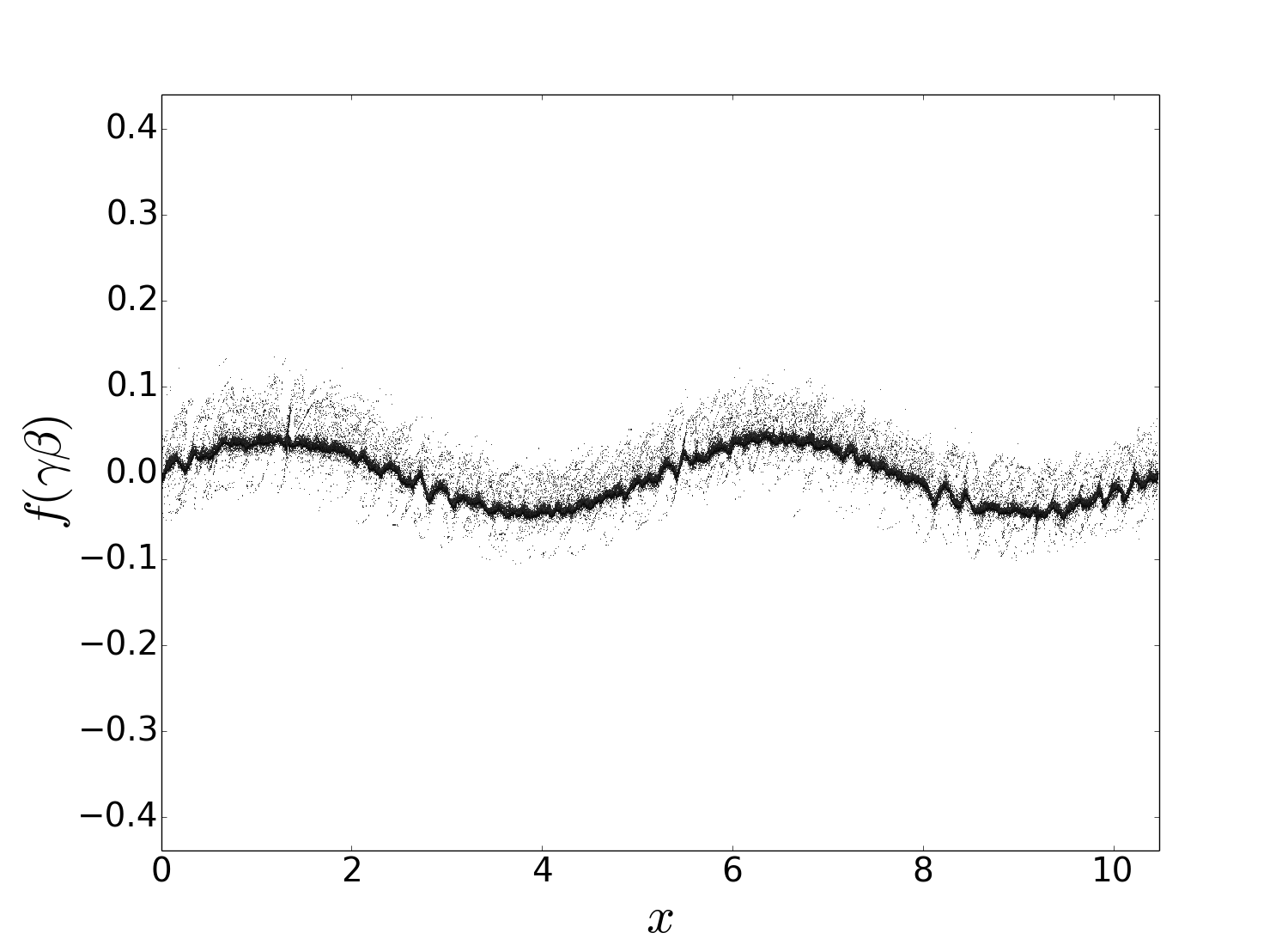}
    \label{2ga_mi1836_Tb10eV_Te01eV_xpx_367wpe_bulk}} 
    & \subfloat[]{\includegraphics[width=0.34\textwidth]{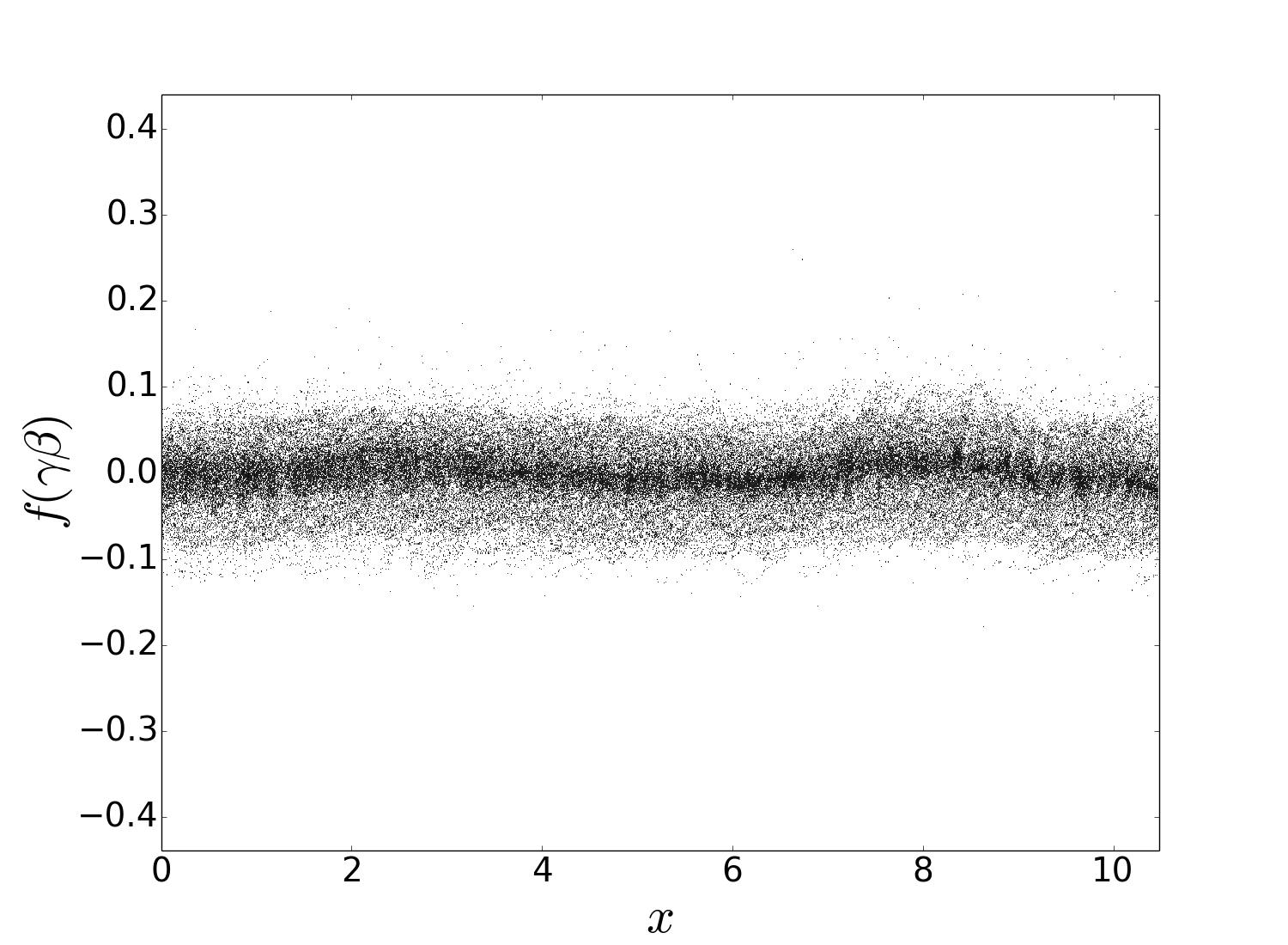}
    \label{2ga_mi1836_Tb10eV_Te01eV_xpx_734wpe_bulk}}\\ 
    \textcolor{red}{ions} & \subfloat[]{\includegraphics[width=0.34\textwidth]{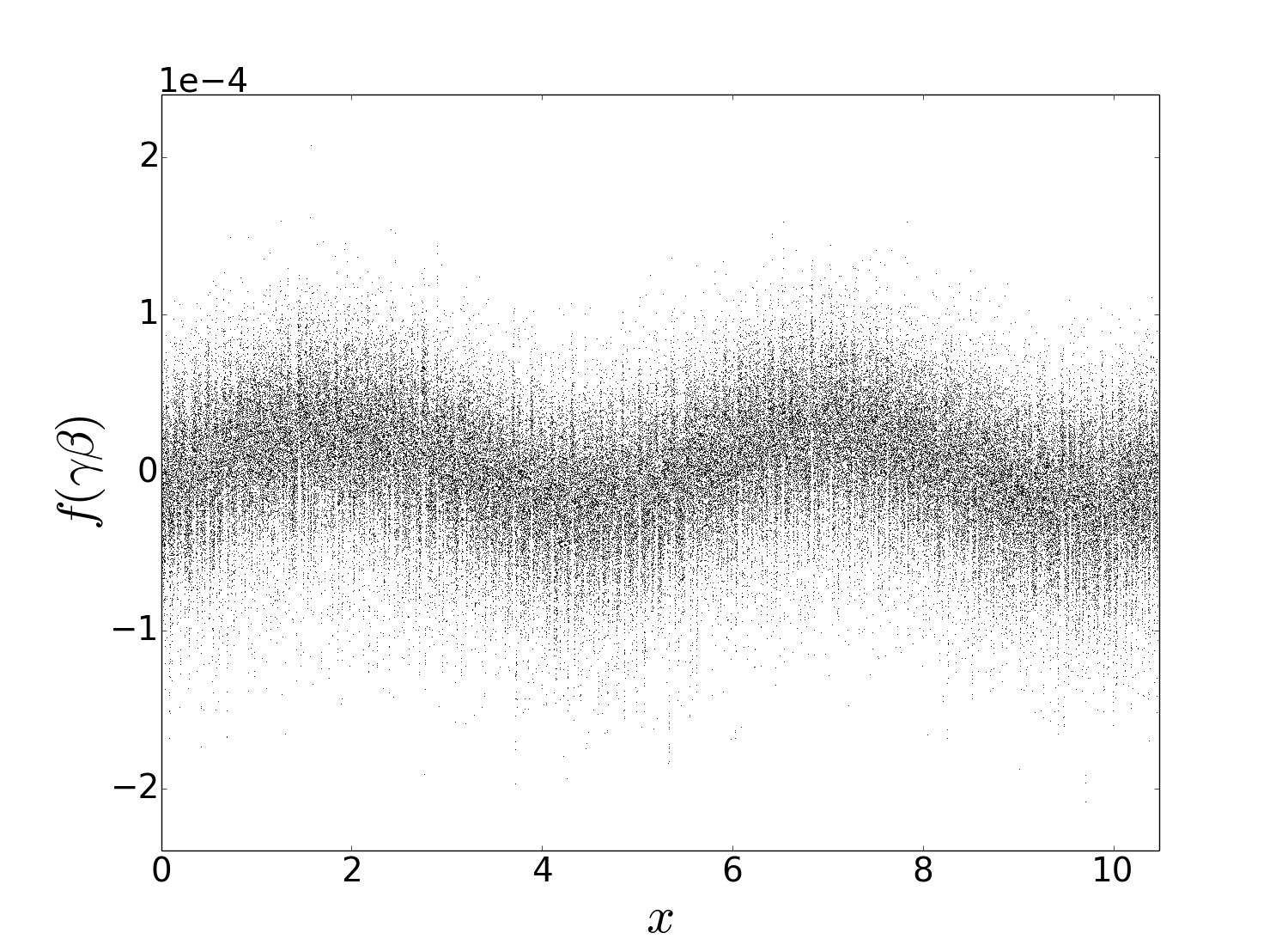}
    \label{2ga_mi1836_Tb10eV_Te01eV_xpx_262wpe_ion}}
    & \subfloat[]{\includegraphics[width=0.34\textwidth]{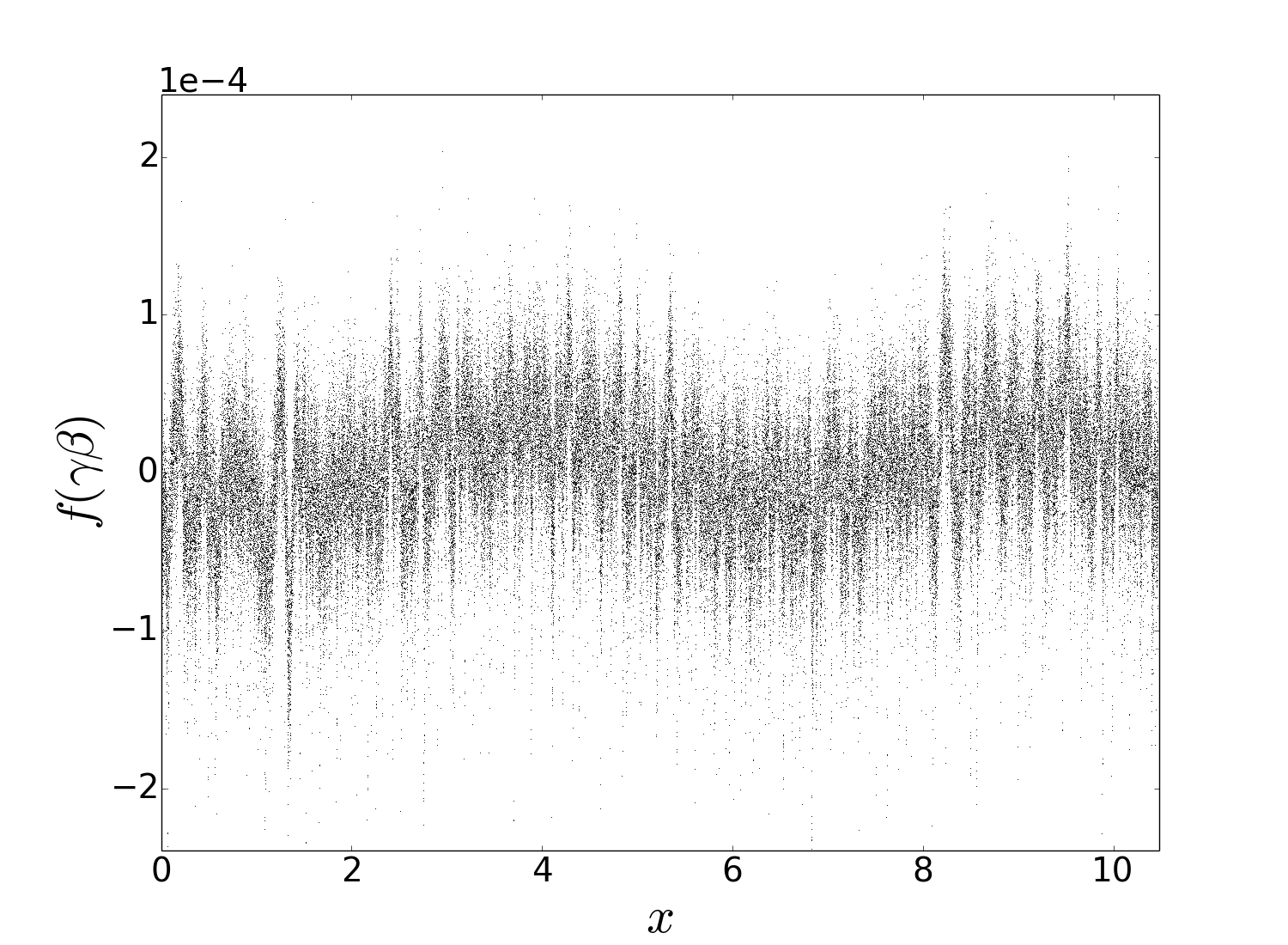}
    \label{2ga_mi1836_Tb10eV_Te01eV_xpx_367wpe_ion}}
    & \subfloat[]{\includegraphics[width=0.34\textwidth]{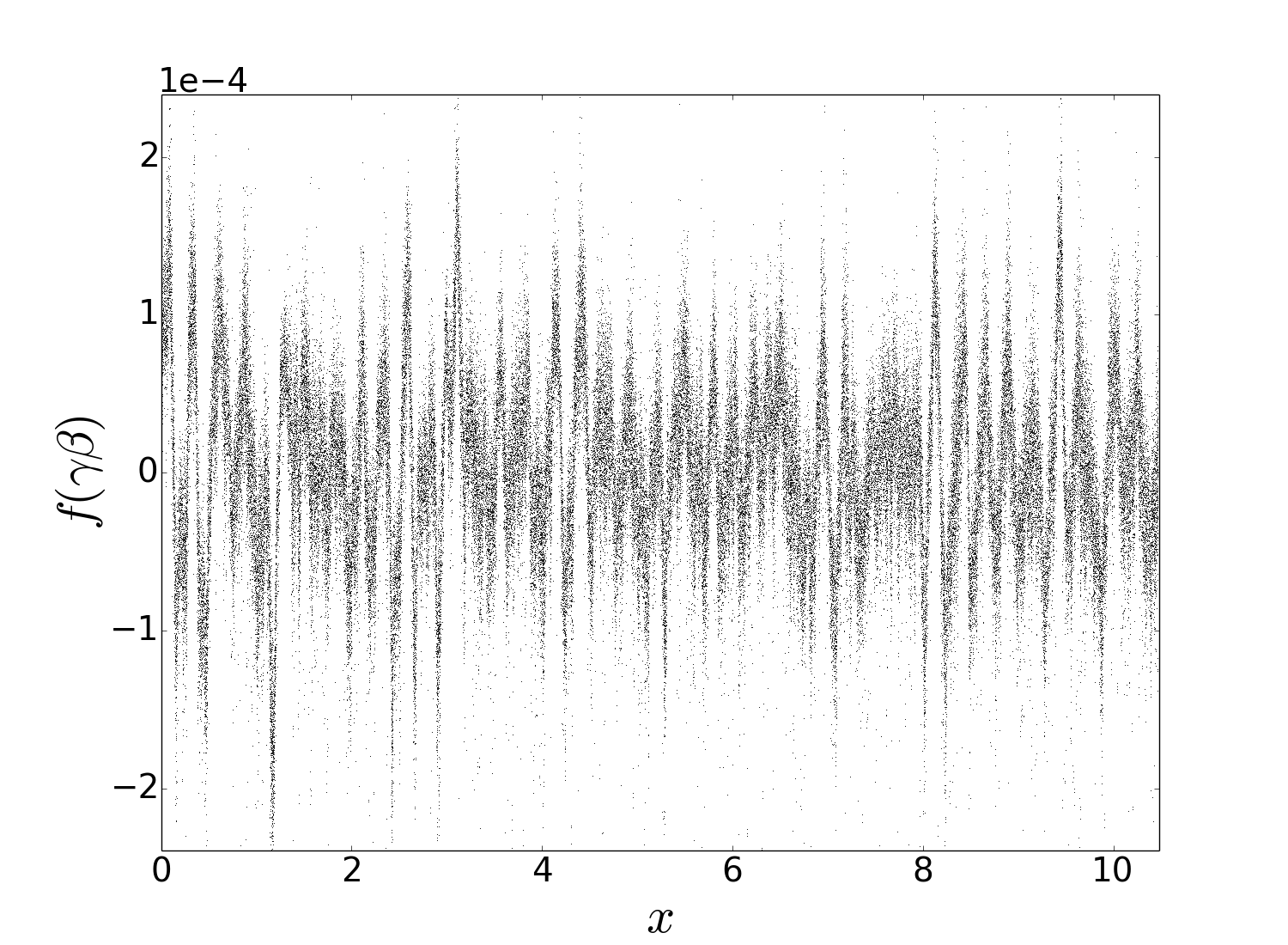}
    \label{2ga_mi1836_Tb10eV_Te01eV_xpx_734wpe_ion}} \\
    & $t= 262 \frac{1}{\omega_{pe}} $ &  $t= 367 \frac{1}{\omega_{pe}} $ & $t= 734 \frac{1}{\omega_{pe}} $ 
\end{tabular}
 \caption{The particle phase space density distributions. The first row shows the distribution of the beam electrons, the second row that of the bulk electrons and the bottow row that of the ions. 
 The first column shows the distributions at the time $t= 262$, the second at the time $t= 367$ and the third one at the time $t= 734$.
 The ion mass is $m_{i}=1836 m_{e}$ and the initial temperatures of the beam and bulk electrons are $T_{b}= 10 eV$ and $T_{e}= 0.1 eV$. \label{phase_space}}
\end{figure*}

We compare in this section the phase space density distributions of the particle species at the times $t=$ 262, 367 and 734, which are marked by the vertical dashed lines in Fig. \ref{density_ener}. 
Figure \ref{phase_space} shows the distributions for the run with $R=$ 1836.
The dilute electron beam drives a Langmuir wave with a positive phase speed $\omega/k$ slightly below the beam speed. This velocity mismatch is caused by the non-zero thermal velocity of the electron beam.
This wave is still growing at the time $t=262$. Its phase is most easily determined from the phase space density distribution of the bulk electrons in Fig. \ref{phase_space}(d), which shows sinusoidal oscillations of the mean speed with $x$. 
The bulk velocity is zero at $x\approx 0.3$ and $x\approx 5.7$ and the electrostatic potential must thus have an extremum at these positions. 
The surrounding electrons are attracted towards these points and the potential is thus positive. 
The beam electrons in Fig. \ref{phase_space}(a) gyrate around the maxima of the positive potential and some are trapped by it. 
The different response of both electron species is caused by a phase speed of the wave, which is much larger in the reference frame of the bulk electrons than in that of the beam electrons. 

The number density of the beam electrons at the cusps at $x=2$ and $x=7.4$ is large and their momentum $\beta \gamma \approx 1$ is below the initial beam momentum, which explains why the beam energy in Fig. \ref{density_ener}(a) has been reduced at this time. 
The distribution of the beam electrons is changing into the phase space vortices or phase space holes, which form when an electrostatic instability with a wave vector that is aligned with the beam velocity \cite{Roberts67,ONeil71} saturates. 
The ions oscillate in the wave's electric field at an amplitude that is small due to the large value of $R$. 

The bulk electrons in Fig. \ref{phase_space}(d) show phase space structures surrounding the dense beam with the wave length $\simeq 3 \cdot 10^{-2}$. These are electron phase space holes that were driven by the Buneman instability. The time, during which such a phase space hole develops, is comparable to the bouncing frequency of a particle with the charge $q$ and mass $m$ in a wave potential $\omega_b={(qkE/m)}^{1/2}$, where $kE$ is the product of wavenumber and electric field amplitude (See Ref. \cite{Dieckmann00} and references cited therein). Electron phase space holes form 5 times faster than ion phase space holes holes even for the lowest ion mass $R=25$. Electron phase space holes are thus responsible for the initial saturation of the Buneman instability in all simulations, which explains why it always saturated at about the same electric field amplitude. 

The electron phase space holes have been driven by the interaction of the charge density distribution of the ion beam with the Langmuir wave, which oscillates at the plasma frequency of the bulk electrons. Their propagation speed in the rest frame of the bulk electrons is close to that of the ions at least for large $R$. In its rest frame, an electron phase space hole is associated with an electrostatic potential that is almost static and it can thus easily accelerate the ions. The speed gain of the ions with $R=1836$ remains small due to the short acceleration time and we observe only small oscillations of the ion velocity at small wavelengths in Fig. \ref{phase_space}(g).

The Langmuir wave, which grew in response to the two-stream instability, has propagated towards increasing x and both cusps in the electron beam distribution have rotated further at the time $t=$ 367. 
The mean momentum of the cusp electrons in Fig. \ref{phase_space}(b) has increased compared to that in Fig. \ref{phase_space}(a), which explains why the beam energy in Fig. \ref{density_ener}(a) has increased. 
This cusp and the current, which is associated with its motion, is causing the periodic exchange of energy between the beam electrons on one hand and the bulk electrons and the electric field on the other hand. 
The beam distribution in Fig. \ref{phase_space}(b) reveals the presence of multiple beams. 
The bouncing in the sinusoidal potential of the electrostatic wave disperses the electrons, which results in a reduction of the amplitude of the energy oscillations. 

The electron phase space holes, which were driven by the Buneman instability, coalesce to larger ones \cite{Roberts67}.
The larger phase space holes in the bulk electrons in Fig. \ref{phase_space}(e) yield now noticable oscillations with the wave length $\simeq 0.1$ in the ion distribution displayed by Fig. \ref{phase_space}(h);
the latter are ion acoustic waves. Ion acoustic waves are linearly undamped if the ratio between the electron temperature and ion temperature is large \cite{Treumann77book}, which is the case in our simulation after the Buneman instability has saturated. We note, however, that the presence of the electron phase space holes means that the ion velocity oscillations may not be linear since electron phase space holes and large ion acoustic waves can couple \cite{Jenab17}. 

Coalescence of the electron phase space holes explains why the characteristic wave number of the waves, which were generated by the Buneman instability, decreases in time in Fig. \ref{t_kx_2ga5_mi1836_T1}. 

The growth of the ion acoustic waves and of the amplitude of the velocity oscillations of the phase space holes in the bulk electrons hints at an instability between both species that is still active long after the Buneman instability saturated. The velocity oscillations of the bulk electrons caused by the two-stream instability have a wave length that is large compared to that of the ion acoustic waves in Fig. \ref{phase_space}(h) and an amplitude that is larger than the electron thermal speed in Fig. \ref{phase_space}(e). The therefrom resulting drift between the bulk electrons and ions is large enough to destabilize the ion acoustic waves and heat the bulk electrons. This mechanism is equivalent to the oscillating two-stream instability \cite{Gupta04} if the laser-generated electrostatic beat wave were replaced by the two-stream mode.

The ion oscillations have increased their amplitude at $t=734$ and the density of the hot component of the bulk electrons has increased. 
The long-wavelength oscillation of the bulk distributions has vanished, which implies that the beam-driven Langmuir wave has been damped out. 
The beam electrons have been dispersed and form now a turbulent distribution with a wide velocity spread in Fig. \ref{phase_space}(c). 
The velocity spread is comparable to the velocity width of the electron phase space holes at the earlier time, which is in turn determined by the velocity interval $v_{tr}={(2qE/mk)}^{1/2}$ around the wave's phase velocity where a particle gets trapped by the wave. 
The beam distribution is well-separated from the bulk electron distribution along the velocity direction.

\begin{figure*}[htbp]
\hspace*{-1.5cm}
\begin{tabular}{*{4}{c}}{\centering}
     & \multicolumn{3}{c}{$m_{i}=25 \: m_{e}$} \\
    \textcolor{blue}{beam} & \subfloat[]{\includegraphics[width=0.34\textwidth]{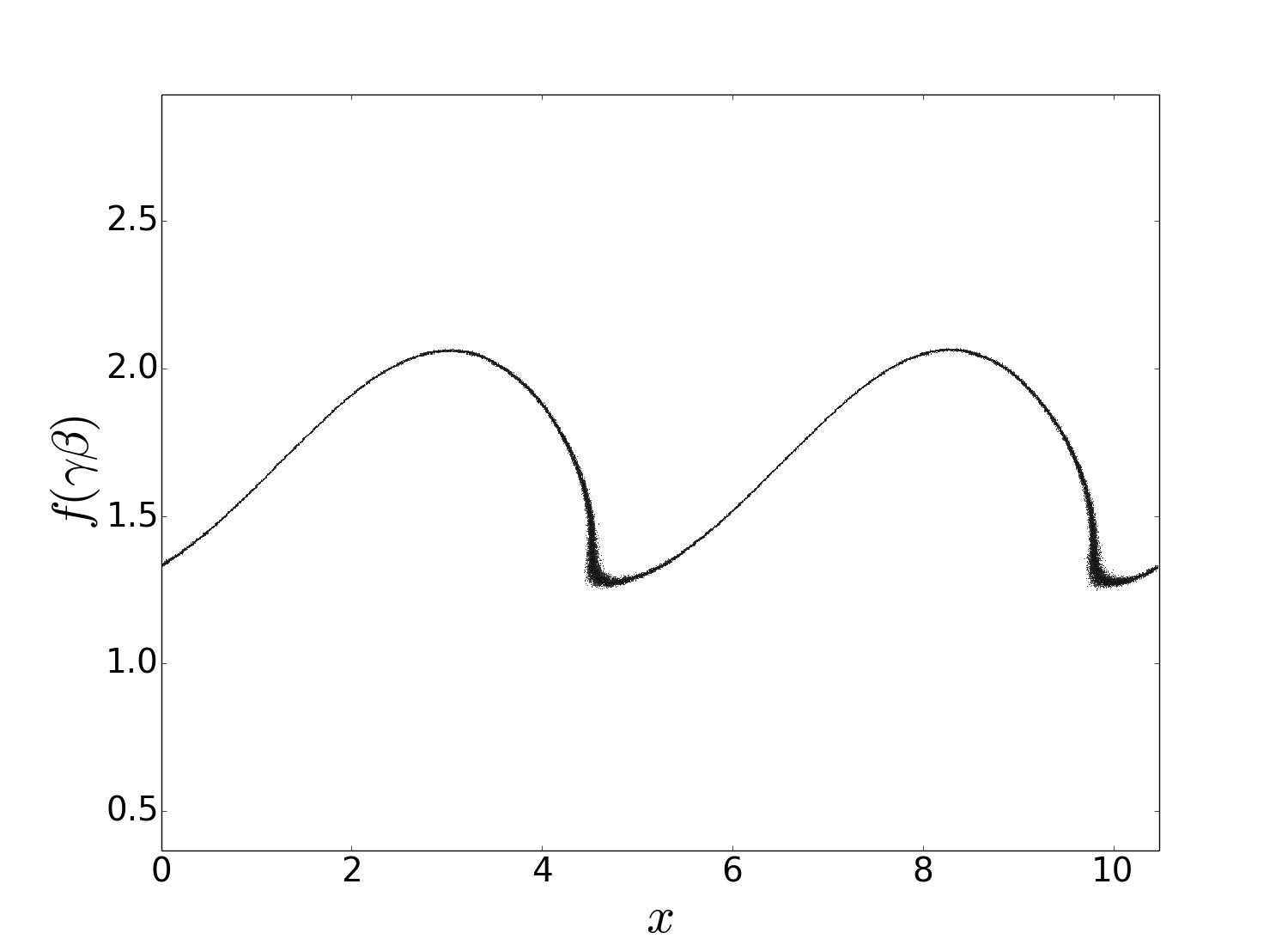}
    \label{2ga_mi25_Tb10eV_Te01eV_xpx_262wpe_beam}} 
    & \subfloat[]{\includegraphics[width=0.34\textwidth]{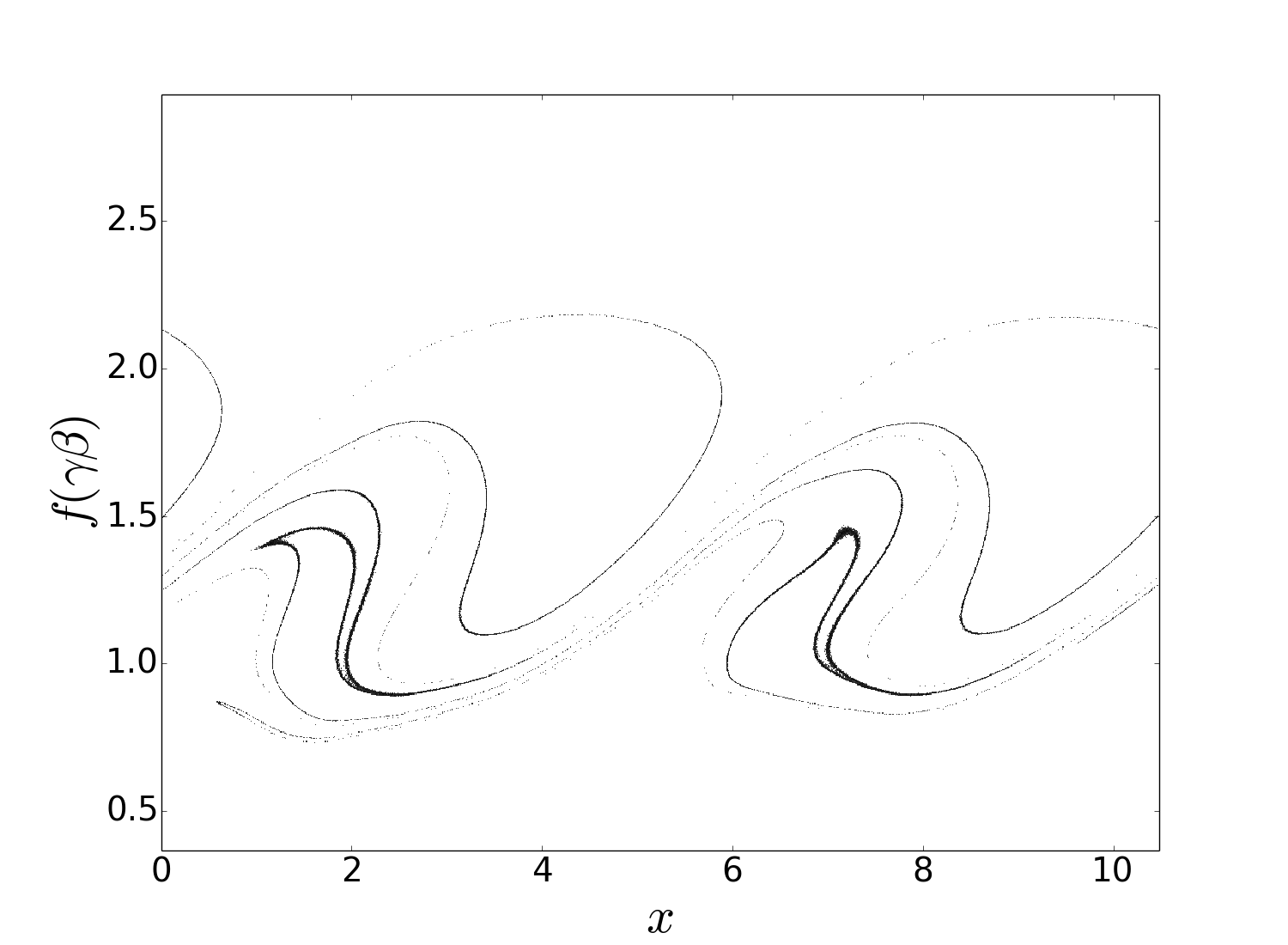}
    \label{2ga_mi25_Tb10eV_Te01eV_xpx_367wpe_beam}} 
    & \subfloat[]{\includegraphics[width=0.34\textwidth]{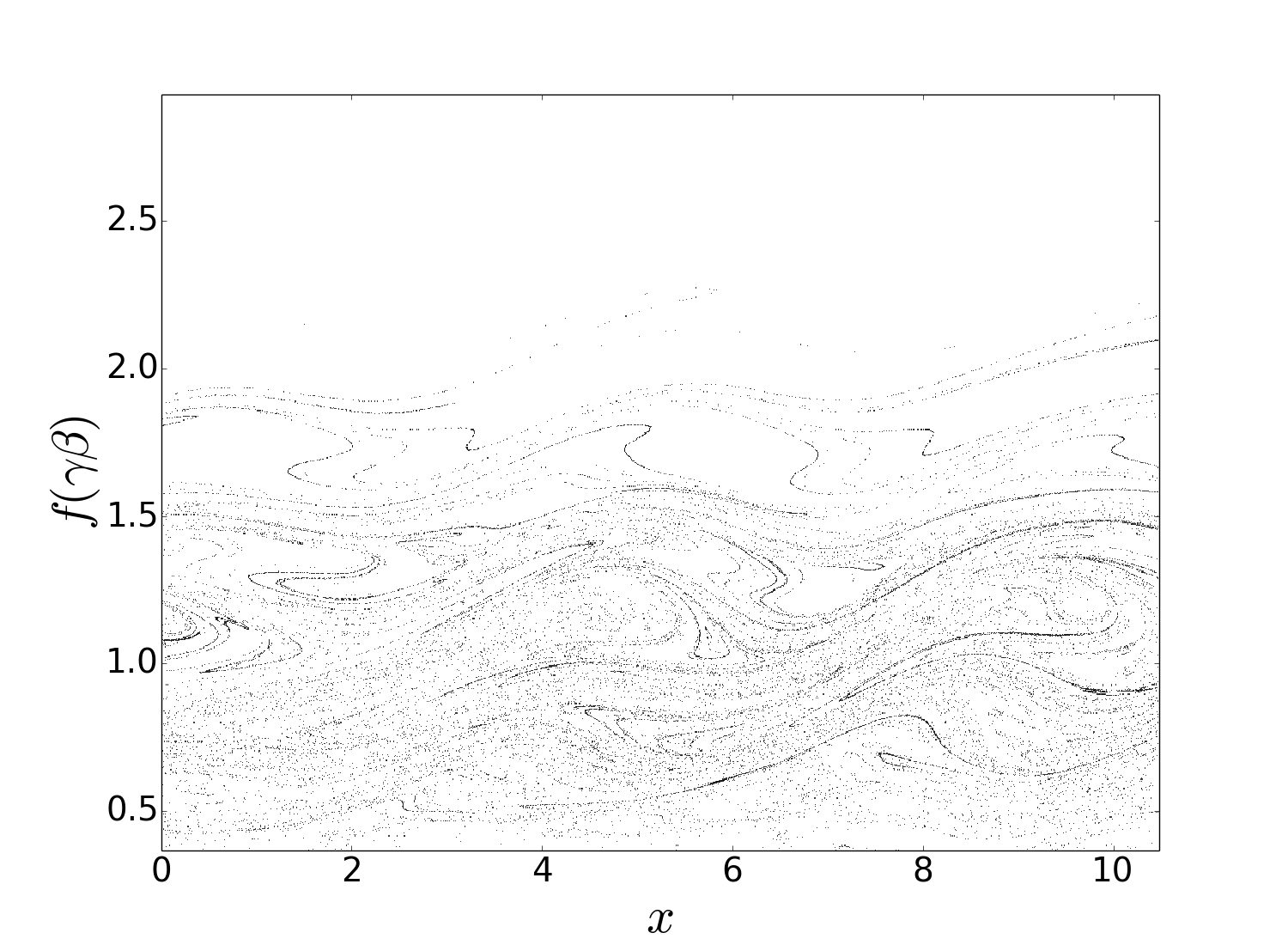}
    \label{2ga_mi25_Tb10eV_Te01eV_xpx_734wpe_beam}} \\
    \textcolor{green}{bulk} & \subfloat[]{\includegraphics[width=0.34\textwidth]{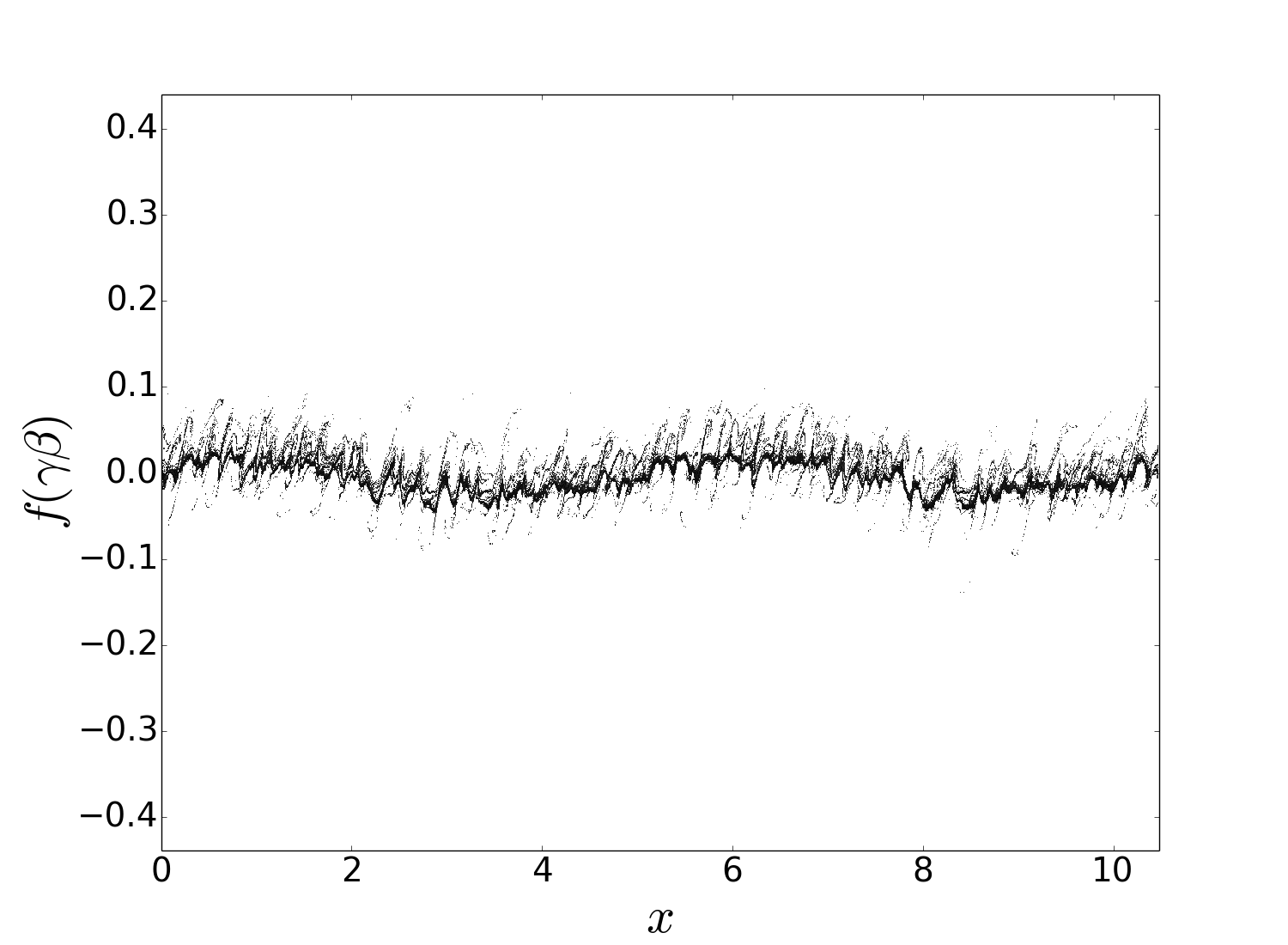}
    \label{2ga_mi25_Tb10eV_Te01eV_xpx_262wpe_bulk}} 
    & \subfloat[]{\includegraphics[width=0.34\textwidth]{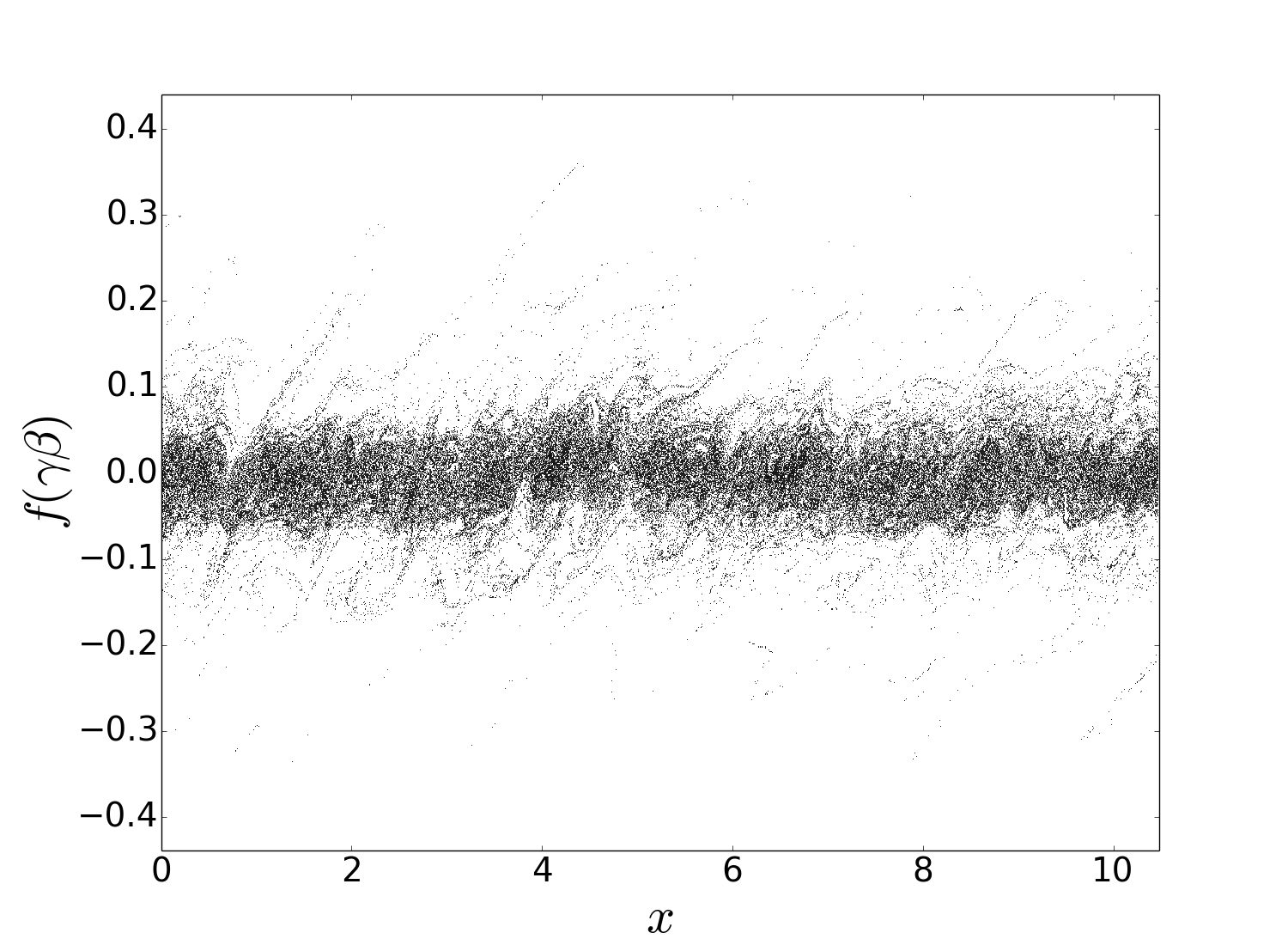}
    \label{2ga_mi25_Tb10eV_Te01eV_xpx_367wpe_bulk}} 
    & \subfloat[]{\includegraphics[width=0.34\textwidth]{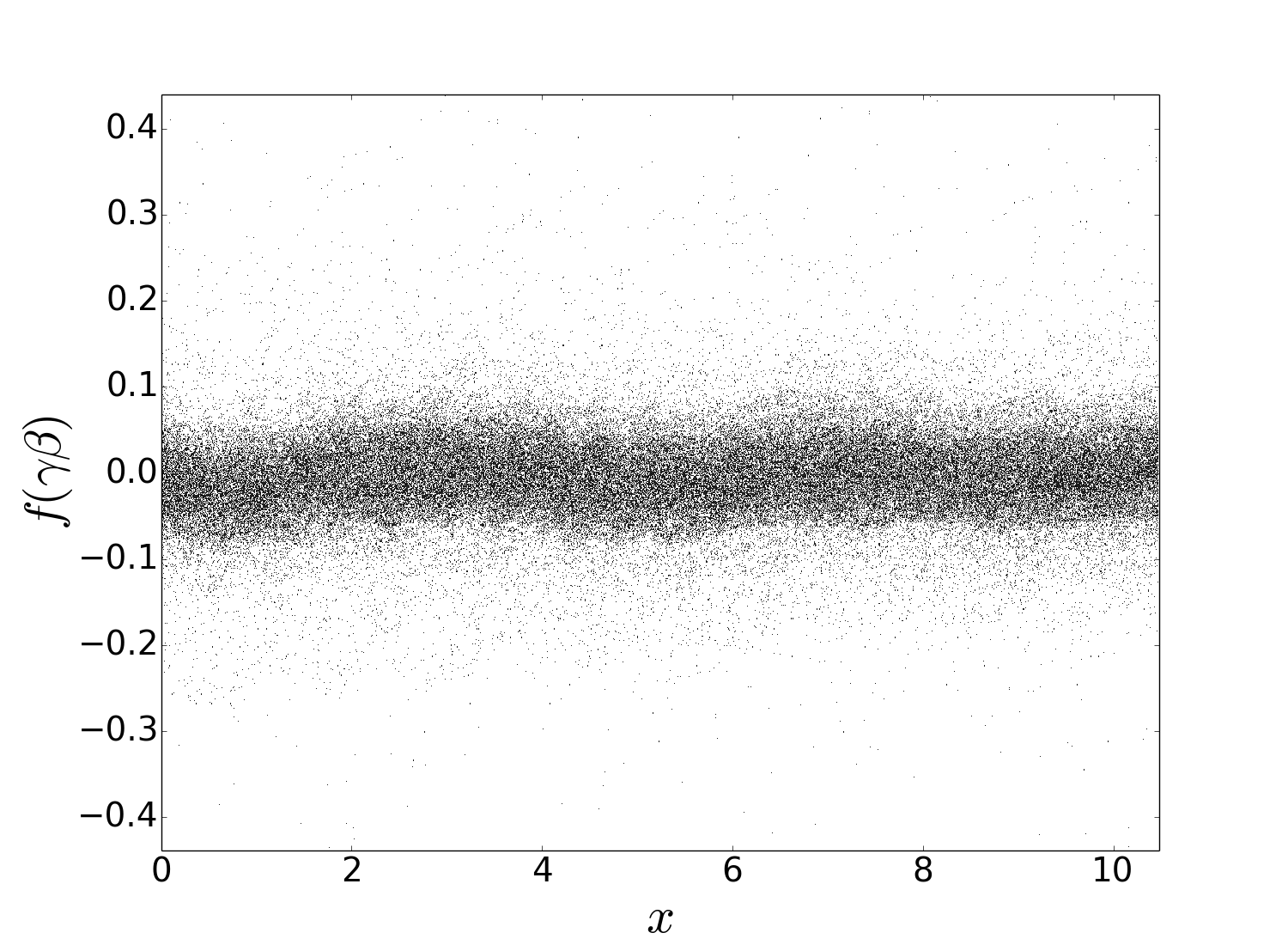}
    \label{2ga_mi25_Tb10eV_Te01eV_xpx_734wpe_bulk}} \\ 
    \textcolor{red}{ions} &\subfloat[]{\includegraphics[width=0.34\textwidth]{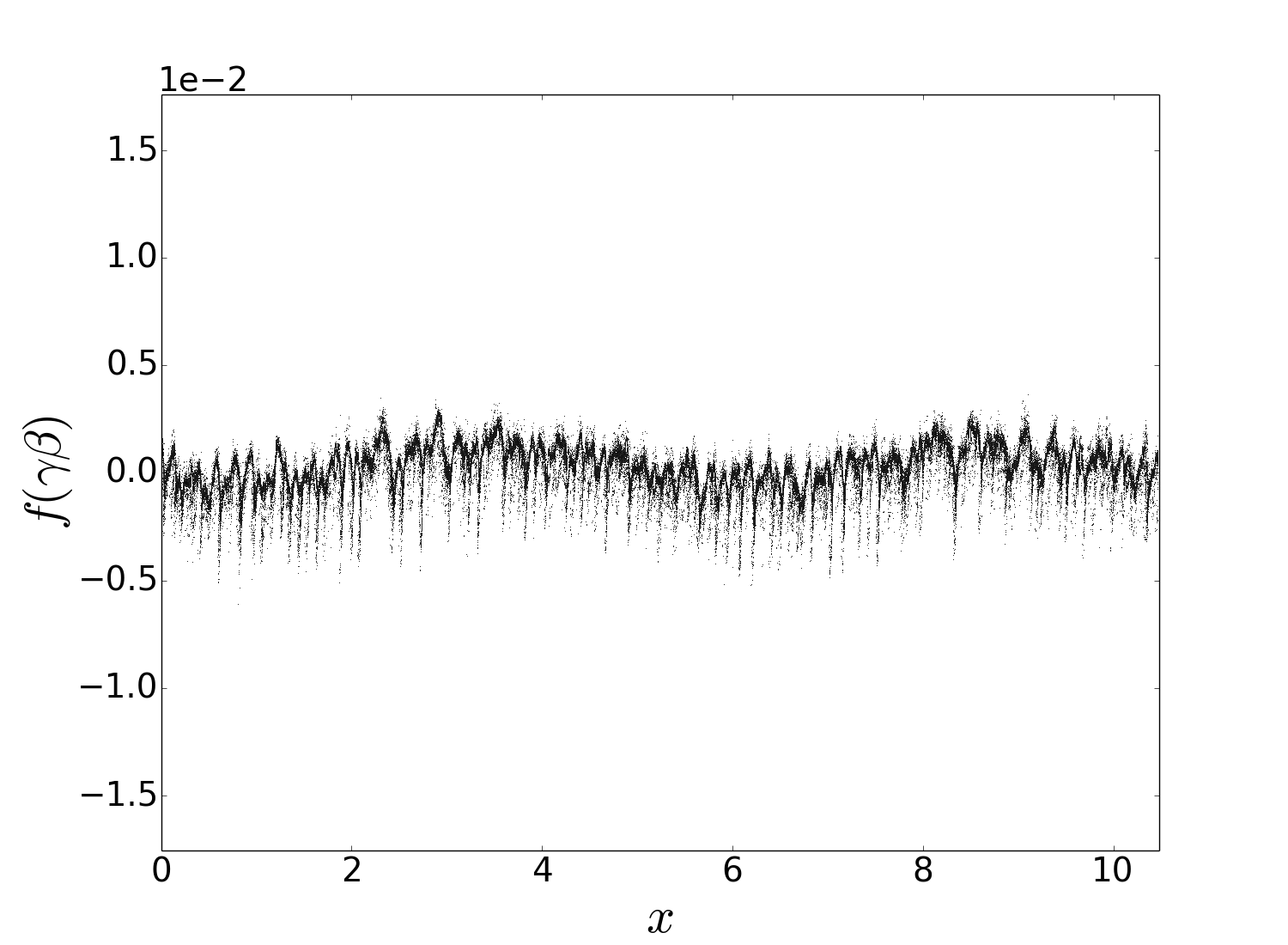}
    \label{2ga_mi25_Tb10eV_Te01eV_xpx_262wpe_ion}}
    & \subfloat[]{\includegraphics[width=0.34\textwidth]{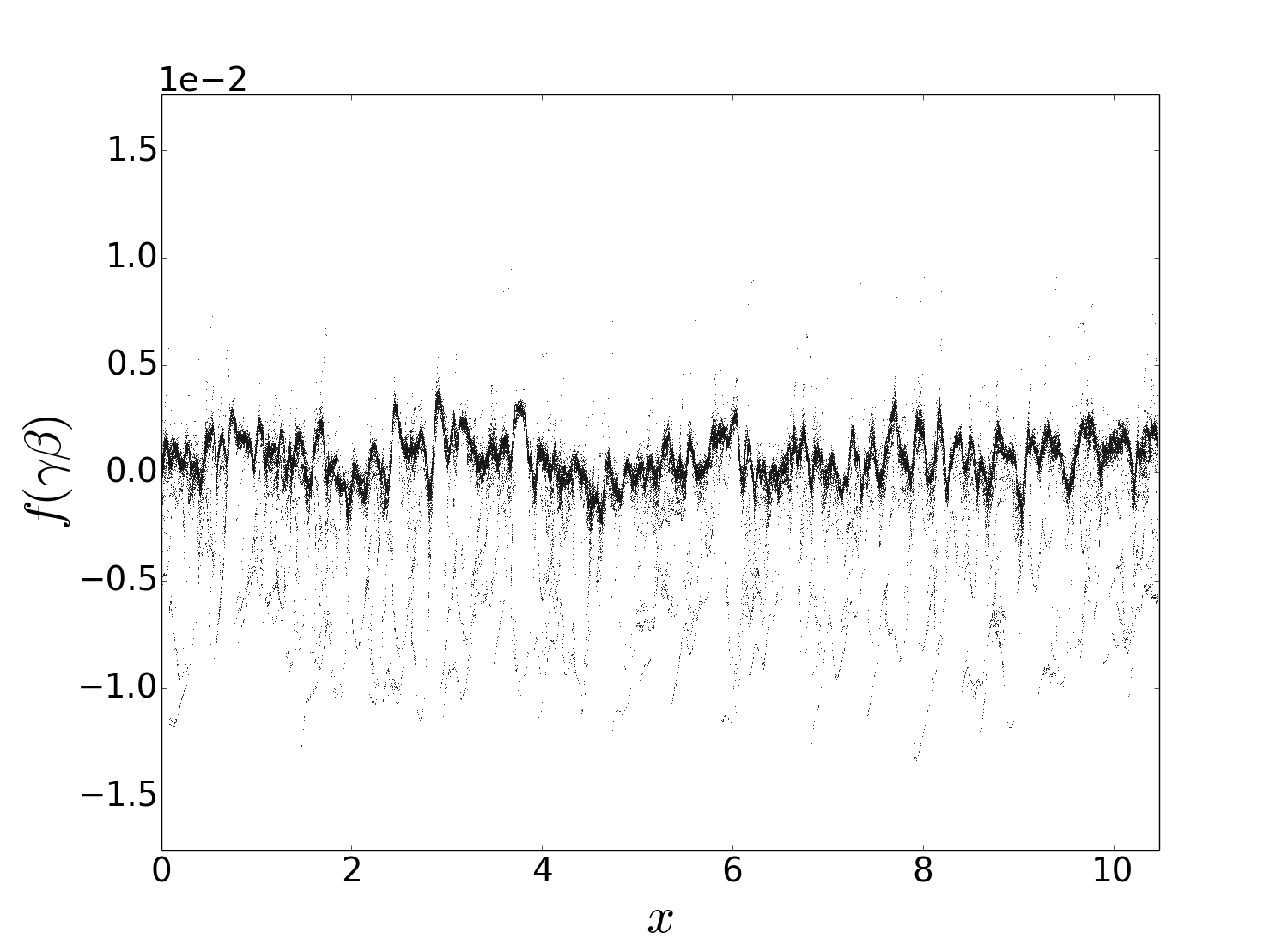}
    \label{2ga_mi25_Tb10eV_Te01eV_xpx_367wpe_ion}}
    & \subfloat[]{\includegraphics[width=0.34\textwidth]{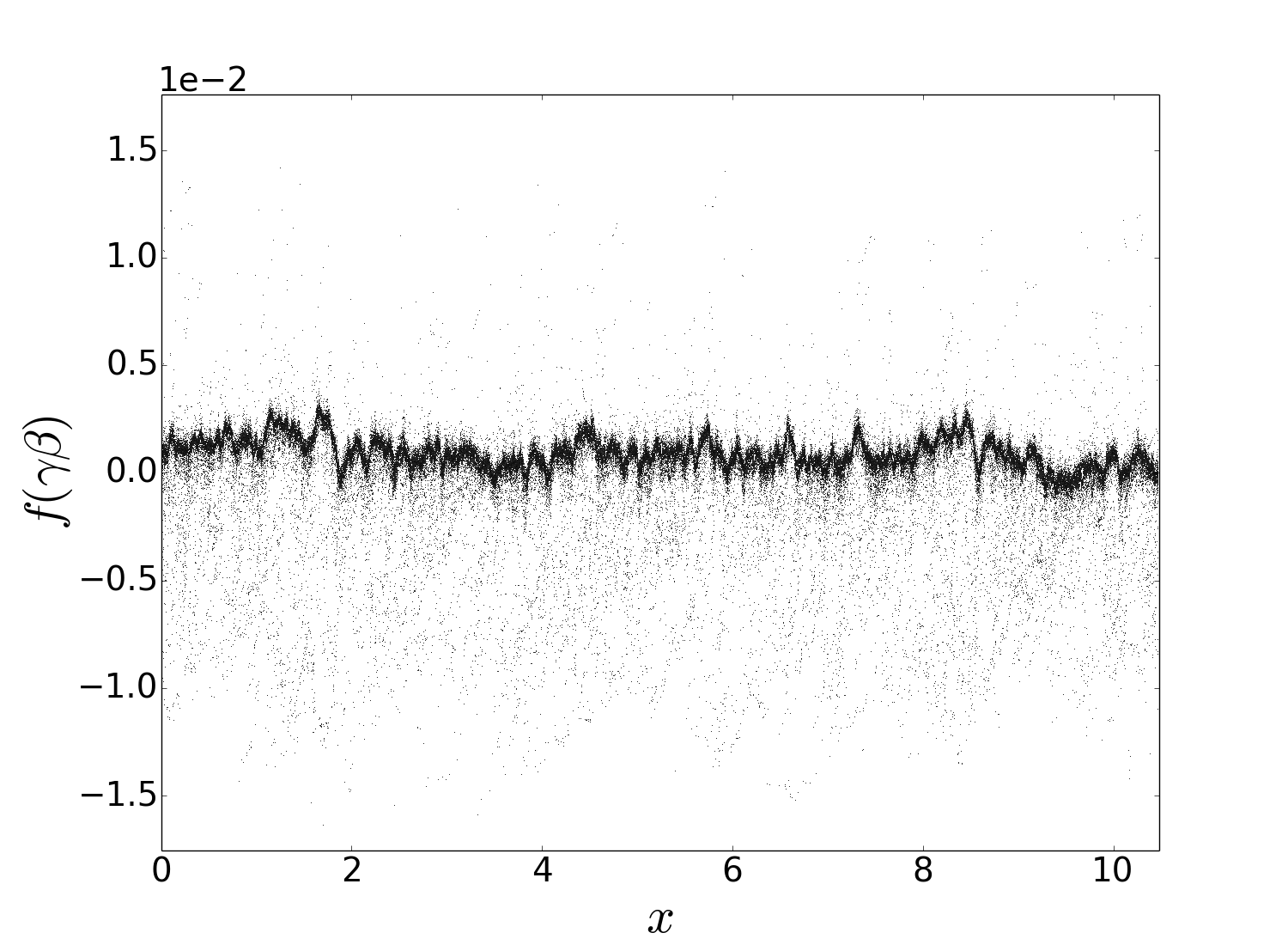}
    \label{2ga_mi25_Tb10eV_Te01eV_xpx_734wpe_ion}} \\
    & $t= 262 \frac{1}{\omega_{pe}} $ &  $t= 367 \frac{1}{\omega_{pe}} $ & $t= 734 \frac{1}{\omega_{pe}} $  
\end{tabular}
 \caption{The particle phase space density distributions. The first row shows the distribution of the beam electrons, the second row that of the bulk electrons and the bottow row that of the ions. 
 The first column shows the distributions at the time $t= 262$, the second at the time $t= 367$ and the third one at the time $t= 734$. 
 The ion mass is $m_{i}=25 m_{e}$ and the initial temperatures of the beam and bulk electrons are $T_{b}= 10 eV$ and $T_{e}= 0.1 eV$. \label{phase_space2}}
\end{figure*}

Figure \ref{phase_space2} shows the phase space density distributions obtained from the run with $R=25$ at the same times as the ones discussed in Fig. \ref{phase_space}. 
The electron beam distribution in Fig. \ref{phase_space2}(a) shows that a large phase space hole is about to form. 
Its shape is practically identical to that in the run with $R=1836$, confirming that the two-stream instability is not affected by the value of $R$.
The earlier development of the Buneman instability and the faster onset of the ion acceleration have perturbed the bulk plasma in Figs. \ref{phase_space2}(d,g) significantly more than that in Figs. \ref{phase_space}(d,g). 
A comparison of Fig. \ref{phase_space2}(h) with Fig. \ref{phase_space}(h) reveals a population of energetic ions for the case $R=25$ that was not present in the simulation with $R=1836$. 

The waves driven by the Buneman instability and the electron phase space holes they drive have a negative phase speed in the simulation frame. 
Ions, which are accelerated in the propagation direction of the electron phase space hole, stay in phase with its electric field for a longer time and are thus accelerated to a larger speed modulus. 
The larger charge-to-mass ratio of ions with $R=25$ meant that they could be accelerated to a larger speed  than those with $R=1836$. 
The ions were accelerated at the expense of the electric field energy. 
The ion acceleration can explain the damping of the Buneman instability-driven waves between $t=262$ and $t=367$ in Fig. \ref{density_ener}(c). 

The electric field, which is associated with the charge density oscillations of the bulk plasma, covers a broad range of wave numbers that extends even to $Z_{TSI}$ (See Fig. \ref{t_kx_2ga5_mi1836_T1}(c)). 
Both bulk distributions still show modulations with a long wavelength, which sustain the wave that grows in response to the two-stream instability. 
Figure \ref{density_ener}(c) shows that the electric field energy decreases after $t=262$, while the energies of the ions and the bulk electrons grow to values well above the ones in the simulations with $R=$ 1836 and 400.
The growth of the energy of the bulk species slows down at $t=367$ and the electric field energy remains high and constant until $t\approx 600$. 
Figure \ref{phase_space2}(b) shows that an electron phase space hole is still present in the electron beam distribution, but it does no longer have the quasi-circular shape as the one in Fig. \ref{phase_space}(b). 
Figures \ref{phase_space2}(e, h) demonstrate that the energy gain of both bulk plasma species is due to a temperature increase and not due to a spatial modulation of the mean speed as in the simulation with $R=1836$. 

The temperature increase of the bulk plasma results in an increase of its thermal pressure. The increasing thermal pressure reduces the density modulation in response to the electric field of the wave. The reduced density modulation causes in turn a reduction of its electric field amplitude and of the spatial modulation of the mean velocity of the bulk electrons and ions. 
The electron phase space hole does no longer interact with the bulk plasma via the electrostatic wave and hence we do no longer observe oscillations due to a periodic energy exchange between the particles and the electric field. 

The ion distribution in Fig. \ref{phase_space2}(h) shows elongated tails that extend to a speed $-10^{-2}$. These tails form during the nonlinear evolution of the Buneman instability. Even though the density of these tails is low, they carry a significant momentum. The mean momentum of the ions has become negative, which is initially compensated for by a momentum increase of the bulk electrons. Electrons gain more speed when they exchange momentum with the ions and they drive a negative total current in the bulk plasma. On average, a positive electric field grows. Its effect can be seen from Fig. \ref{phase_space2}(c). The electron beam distribution is as diffuse as that in Fig. \ref{phase_space}(c) and its velocity width is comparable. The mean velocity of the beam is, however, lower by a value 0.3 for all x. The lost kinetic energy is transferred to the bulk plasma. This process has been discussed previously in Ref. \cite{Lovelace71}.

\section{Discussion}

We have studied the interplay of the Buneman instability with the two-stream instability by solving the linear dispersion relation and by means of particle-in-cell simulations. The initial conditions for the instabilities were selected such that the Buneman instability would grow as fast as the two-stream instability for the {ion-to-electron mass} ratio of protons and faster for a reduced ion mass. This case study is a testbed for whether or not the plasma evolution is determined by the instability with the largest exponential growth rate. That criterion was used by Ref. \cite{Bret10b} to investigate the impact a reduced mass ratio has on the evolution of a relativistic electron beam that interacts with background electrons and ions. 

Our results are as follows. Both instabilities grew independently and at the expected exponential growth rate during their initial growth phase. The wave, which was driven by the Buneman instability, saturated at an amplitude well below that of the two-stream instability. Variations of the exponential growth rate of the Buneman instability with the ion-to-electron mass $R$ had no consequence for the plasma evolution because it always saturated first for the initial conditions we considered. 

The criterion for the most important instability in a beam-plasma system used by Ref. \cite{Bret10a} assumes that the instability saturates at a time, which is proportional to its inverse exponential growth rate. This time scales as $\propto R^{-1/3}$ for the Buneman instability. The electrostatic waves in our simulations started to grow only after a significant delay in time and the growth time, measured from the onset of the wave growth, scaled like $R^{-1/2}$. The discrepancy is large especially for low values of $R$. 

Artifacts introduced by a reduced value for $R$ were more pronounced during the non-linear evolution phase of the Buneman instability. The Buneman instability saturated for all $R$ by trapping electrons from the bulk distribution. Merging of the electron phase space holes, which formed during the saturation of the instability, shifted the characteristic wave number of the electrostatic oscillations to lower wave numbers. Eventually they started to interact with the waves driven by the two-stream instability and electrostatic oscillations grew over a broad frequency band. The amplitudes of these oscillations relative to those of the two-stream mode increased with decreasing $R$.

The electrostatic wave, which was driven by the two-stream instability, grew to a larger amplitude when $R$ was large and it was stable over a longer time. We observed for $R=1836$ a periodic exchange of energy between the two-stream mode on one hand and the bulk plasma and the electric field on the other hand. These oscillations are caused by the well-known trapping of beam electrons. These oscillations were damped for $R=400$ and asymptotically damped for $R=25$. Decreasing $R$ resulted in a faster and more pronounced energy loss of the beam electrons to the bulk electrons and ions. In particular the ion energy increased for $R=25$.

A reduction of $R$ below 100 resulted in a doubling of the energy loss of the beam electrons and in a drastic reduction of their mean speed. Simulations that address the propagation of relativistic, cold and dilute electron beams through a background plasma should thus keep a value of $R$ close to the correct one.

\section{ACKNOWLEDGMENTS}

This work was supported by the French National Research Agency Grant ANR-14-CE33-0019 MACH. This work was also granted access to the HPC resources of CINES and TGCC under allocations A0020510052 and A0030506129 made by GENCI (Grand Equipement National de Calcul Intensif), and has been partially supported by the 2015-2019 grant of the Institut Universitaire de France obtained by CELIA.


\begin{thebibliography}{}
\bibitem{Lovelace71} R. V. Lovelace, and R. N. Sudan, Phys. Rev. Lett. \textbf{27}, 1256 (1971). 
\bibitem{Thode75} L. E. Thode, and R. N. Sudan, Phys. Fluids \textbf{18}, 1564 (1975). 
\bibitem{Buneman58} O. Buneman, Phys. Rev. Lett. \textbf{1}, 8 (1958).
\bibitem{Treumann77book} R.  A. Treumann and  W. Baumjohann, Advanced  Space  Plasma  Physics,  Imperial   College Press, London, 1977
\bibitem{Califano98} F. Califano, R. Prandi, F. Pegoraro, and S. V. Bulanov, Phys. Rev. E \textbf{58}, 7837 (1998). 
\bibitem{Bret10a} A. Bret, L. Gremillet, and M. E. Dieckmann, Phys. Plasmas \textbf{17}, 120501 (2010).
\bibitem{Bret10b} A. Bret, and M. E. Dieckmann, Phys. Plasmas \textbf{17}, 032109 (2010). 
\bibitem{Kazimura98} Y. Kazimura, J. I. Sakai, T. Neubert, and S. V. Bulanov, Astrophys. J. \textbf{498}, L 183 (1998).
\bibitem{Honda00} M. Honda, J. Meyer-ter-Vehn, and A. Pukhov, Phys. Plasmas \textbf{7}, 1302 (2000).
\bibitem{Silva03} L. O. Silva, R. A. Fonseca, J. W. Tonge, J. M. Dawson, W. B. Mori, and M. V. Medvedev, Astrophys. J. \textbf{596}, L121 (2003).
\bibitem{Sakai04} J. I. Sakai, R. Schlickeiser, and P. K. Shukla, Phys. Lett. A \textbf{330}, 384
\bibitem{Jaroschek05} C. H. Jaroschek, H. Lesch, and R. A. Treumann, Astrophys. J. \textbf{618}, 822 (2005).
\bibitem{Nishikawa09} K. I. Nishikawa, J. Niemiec, P. E. Hardee, M. Medvedev, H. Sol, Y. Mizuno, B. Zhang, M. Pohl, M. Oka, and D. H. Hartmann, Astrophys. J. \textbf{698}, L10 (2009).
\bibitem{Bret13} A. Bret, A. Stockem, F. Fiuza, C. Ruyer, L. Gremillet, R. Narayan, and L. O. Silva, Phys. Plasmas \textbf{20}, 042102 (2013).
\bibitem{Thode73} L. E. Thode, and R. N Sudan, Phys. Rev. Lett. \textbf{30}, 732 (1973).
\bibitem{Dieckmann00} M. E. Dieckmann, P. Ljung, A. Ynnerman, and K. G. McClements, Phys. Plasmas \textbf{7}, 5171 (2000).
\bibitem{Muschietti90} L. Muschietti, Solar Phys. \textbf{130}, 201 (1990).
\bibitem{Kontar01} E. P. Kontar, Astron. Astrophys. \textbf{375}, 629 (2001).
\bibitem{Klein05} K. L. Klein, S. Krucker, G. Trottet, and S. Hoang, Astron. Astrophys. \textbf{431}, 1047 (2005).
\bibitem{Hamish13} H. A. S. Reid, and E. P. Kontar, Astrophys. J. \textbf{721}, 864 (2010).
\bibitem{Hamish14} H. A. S. Reid, and H. Ratcliffe, Res. Astron. Astrophys. \textbf{14}, 773 (2014).
\bibitem{Tabak} M. Tabak, J. Hammer, M. E. Glinsky, W. L. Kruer, S. C. Wilks, J. Woodworth, E. M. Campbell, and M. D. Perry, Phys. Plasma \textbf{1}, 1629 (1994).
\bibitem{Esarey09} E. Esarey, C. B. Schroeder, and W. P. Leemans, Rev. Mod. Phys. \textbf{81}, 1229 (2009).
\bibitem{Sarri17} J. Warwick, T. Dzelzainis, M. E. Dieckmann, W. Schumaker, D. Doria, L. Romagnani, K. Poder, J. M. Cole, A. Alejo, M. Yeung, K. Krushelnick, S. P. D. Mangles, Z. Najmudin, B. Reville, G. M. Samarin, D. D. Symes, A. G. R. Thomas, M. Borghesi, and G. Sarri, Phys. Rev. Lett. \textbf{119}, 185002 (2017).
\bibitem{Sironi14} L. Sironi, and D. Giannios, Astrophys. J. \textbf{787}, 49 (2014).
\bibitem{Tzoufras06} M. Tzoufras, C. Ren, F. S. Tsung, J. W. Tonge, W. B. Mori, M. Fiore, R. A. Fonseca, and L. O. Silva, Phys. Rev. Lett. \textbf{96}, 105002 (2006).
\bibitem{EPOCH} T. D. Arber, K. Bennett, C. S. Brady, A. Lawrence-Douglas, M. G. Ramsay, N. J. Sircombe, P. Gillies, R. G. Evans, H. Schmitz, A. R. Bell, and C. P. Ridgers, Plasma Phys. Controll. Fusion \textbf{57}, 113001 (2015).
\bibitem{Esirkepov00CPC} T.Zh. Esirkepov, Comput. Phys. Commun. \textbf{135}, 144 (2001).
\bibitem{Roberts67} K. V. Roberts, and H. L. Berk, Phys. Rev. Lett. \textbf{19}, 297 (1967).
\bibitem{ONeil71} T. M. O'Neil, J. H. Winfrey, and J. H. Malmberg, Phys. Fluids \textbf{14}, 1204 (1971).
\bibitem{Jenab17} S. M. Hosseini Jenab, and F. Spanier, IEEE Trans. Plasma Sci. \textbf{45}, 2182 (2017).
\bibitem{Gupta04} D. N. Gupta, K. P. Singh, A. K. Sharma, and N. K. Jaiman, Phys. Plasmas \textbf{11}, 5250 (2004). 
\end{thebibliography}
\end{document}